\documentclass[lettersize,journal]{IEEEtran}
\usepackage{amsmath,amsfonts}
\usepackage{algorithmic}
\usepackage{algorithm}
\usepackage{array}
\usepackage[caption=false,font=normalsize,labelfont=sf,textfont=sf]{subfig}
\usepackage{textcomp}
\usepackage{stfloats}
\usepackage{url}
\usepackage{verbatim}
\usepackage{graphicx}
\usepackage{cite}
\usepackage{caption}
\usepackage{multirow}
\usepackage[numbers,sort&compress]{natbib}
\usepackage[table,xcdraw]{xcolor}

\begin{document}

\title{Towards AI-Native RAN: An Operator’s Perspective of \\ 6G Day 1 Standardization}

\author{Nan Li, Qi Sun, Lehan Wang, Xiaofei Xu, Jinri Huang, Chunhui Liu, Jing Gao, \\Yuhong Huang, and Chih-Lin I,~\IEEEmembership{Life Fellow,~IEEE}
\thanks{Nan Li (corresponding author) is with Beijing University of Posts and Telecommunications, China, and also with China Mobile Research Institute, China; Nan Li, Qi Sun, Lehan Wang, Xiaofei Xu, Jinri Huang, Chunhui Liu, Jing Gao and Chih-Lin I are with China Mobile Research Institute, China; Yuhong Huang is with China Mobile Research Institute, China, and also with ZGC Institute of Ubiquitous-X innovation and Applications, China.}
}



\maketitle

\begin{abstract}

Artificial Intelligence/Machine Learning (AI/ML) has become the most certain and prominent feature of 6G mobile networks. Unlike 5G, where AI/ML was not natively integrated but rather an add-on feature over existing architecture, 6G shall incorporate AI from the onset to address its complexity and support ubiquitous AI applications. Based on our extensive mobile network operation and standardization experience from 2G to 5G, this paper explores the design and standardization principles of AI-Native radio access networks (RAN) for 6G, with a particular focus on its critical Day 1 architecture, functionalities and capabilities. We investigate the framework of AI-Native RAN and present its three essential capabilities to shed some light on the standardization direction; namely, AI-driven RAN processing/optimization/automation, reliable AI lifecycle management (LCM), and AI-as-a-Service (AIaaS) provisioning. The standardization of AI-Native RAN, in particular the Day 1 features, including an AI-Native 6G RAN architecture, were proposed. For validation, a large-scale field trial with over 5000 5G-A base stations have been built and delivered significant improvements in average air interface latency, root cause identification, and network energy consumption with the proposed architecture and the supporting AI functions. This paper aims to provide a Day 1 framework for 6G AI-Native RAN standardization design, balancing technical innovation with practical deployment.

\end{abstract}


\section{Introduction}

Artificial Intelligence (AI) technology has made unprecedented progress in recent years, giving rise to novel service applications that may fundamentally change the way our societies produce and consume. AI technology is also being retrofitted into the mobile networks. In the past few years, the telecom industry has gradually recognized its potential in enhancing network performance, optimizing user experience, reducing power consumption, and automating network management. Following the endeavors in the core network and the management domain \cite{3gpp.23.288,3gpp.28.104,3gpp.28.105}, additional support for AI in virtually all protocol layers is being introduced to the RAN, which largely determines the performance of the mobile networks \cite{3gpp.37.817,3gpp.38.843,3gpp.38.744}. However, the AI functionality introduced in 5G has largely followed an add-on paradigm, which has constrained the potential of AI by legacy architecture, pre-existing protocols, and limited system flexibility. 

The deep convergence of RAN and AI is a cornerstone for the evolution towards 6G \cite{itu}. 
First, AI is promising in addressing some of the most sophisticated challenges in 6G RAN. For example, it may enable low-complexity signal processing for massive Multiple-Input Multiple-Output (MIMO), a technology that has been used to combat the adverse and time-varying radio channels, whose computational complexity rises steeply as hundreds of antenna elements will be employed in a 6G radio unit. Another example is the radio resources allocation and scheduling tasks in complex network topologies with massive User Equipments (UEs). Those tasks typically involve large-scale and non-linear problems for which real-time algorithms have to compromise performance, or even problems that are difficult to model mathematically. Moreover, the distributed RAN sites equipped with AI capabilities are envisioned to serve as the key enablers for the pervasive AI applications, which would demand both low latency and high compute. To that end, the 6G RAN must be designed with AI in a native approach. This includes both AI for RAN—leveraging AI to enhance RAN performance, adaptability, and efficiency—and RAN for AI, which provides beyond communication support to enable pervasive, high-performance AI applications across the network. By unlocking new levels of service capability, energy efficiency, operational intelligence, and cost-effectiveness, AI-Native RAN will serve as a cornerstone of 6G—and ultimately determine its ability to succeed at scale and deliver on its transformative promise.

Though the concept of AI-Native RAN for 6G has gained traction in the telecom industry, a general consensus on its definition and its pathway is yet to be achieved. Various interpretations can be seen from the stakeholders with diverse perspectives. To list a few, Ref.~\cite{wu2021toward} highlights a converged communication and computing architecture that enables dynamic AI service deployment and task-oriented connectivity. 
Ref.~\cite{iovene2023defining} interprets AI-Native concept as the pervasive AI capabilities design, deployment, operation, and maintenance. 
Ref.~\cite{hoydis2021} is focused on the air interface and proposes AI/ML integration for dynamic adaptation to hardware constraints, radio environments, and application requirements. Ref.~\cite{QC3GPPAI} discusses several possible interpretations from a standardization angle, such as the AI/ML availability and flexibility for AI/ML based implementations. Ref.~\cite{kundu2025ai} decomposes the concept into three aspects, namely “AI for RAN”, “AI and RAN” and “AI on RAN”. Refs.~\cite{sun2023intelligent,li2025} claim AI/ML as the foundation for RAN evolution, enabling hierarchical and distributed intelligence across the network. Without a clear and shared understanding of the scope and goals of AI-Native RAN, standards development organizations (SDOs) may suffer from ambiguities, misalignment, or even conflicting efforts in their 6G standardization.

In the process of 6G RAN standardization, the Day 1 standards have special significance against the later releases. Failure to address the AI-Native concept in initial standards could perpetuate the ‘AI-as-afterthought’ limitations we faced in 5G, and eventually undermine the commercial viability of AI-Native RAN in 6G. It should also be noted that the practical needs and concerns of mobile network operators (MNOs) should be carefully considered and reflected in standardization efforts.The MNOs have been confronted with many difficulties, including inefficient utilization of 5G infrastructure, limited sources of revenue, slow innovations, etc. It is desired that novel architectural and functional designs in 6G AI-Native RAN can help to tackle those challenges and ensure a healthy and sustainable telecom ecosystem.

This paper aims to investigate the connotations and impacts of AI-Native RAN from the perspective of a MNO. The main contributions of this paper is as follows. First, We investigates the essential capabilities of AI-Native RAN. Furthermore, Day 1 standardization principles, design considerations and the essential features are presented with an AI-Native RAN architecture proposed, drawing from 4G/5G standardization experience and commercial deployment insights. The potential collaboration between 3GPP and other standard development organizations (SDOs) are also investigated. These guidelines aim to ensure that AI-Native RAN capabilities can be fully realized from the first release of 6G and support their sustainable evolution within an open ecosystem.
Finally, we share a national wide, large-scale effort involving over 5000 base stations in 5G-Advanced (5G-A) networks as an early exploration toward AI-Native design. The field trials validates the key architectural features and three typical use cases, offering valuable practical insights for the path toward a fully realized AI-Native RAN in 6G.

This paper is structured as follows. A brief review of the 5G/5G-A progress in the AI/ML related work is conducted in Section II, along with remaining issues for 6G, and in particular for 6G RAN. The concept, essential capabilities and day 1 standardization considerations of AI-Native RAN are analyzed in details in Section III. The overall standardization considerations, key designs on architecture and building blocks, and the collaboration among SDOs are presented in Section IV. In Section V, the large-scale field trials by China Mobile are discussed. The conclusion of this paper is given in Section V.

\section{Review of AI/ML integration in 5G/5G-A}
The pivot role of AI/ML for RAN evolution has become prominent in the 5G timeframe, with notable progress achieved in several SDOs.

In 3GPP, the standardization efforts on AI/ML highlight a comprehensive approach, despite constrained the existing 5G architecture. While the service and system aspects (SA) work groups (WGs) engage in the relevant activities on various aspects (i.e., service requirements, core network, management, security, and application enablement, see \cite{3gpp.23.288,3gpp.28.104,3gpp.28.105, 3gpp.22.874,3gpp.23.501,3gpp.26.531,3gpp.26.927}), the RAN WGs focus on the RAN-specific AI/ML use cases, functionalities, and the associated LCM over the air interface and the Xn interface. The AI/ML enabled use cases include channel state information (CSI) feedback, beam management, positioning, mobility, network energy saving and load balancing, with the promise to enhance the air interface efficiency, enable proactive mobility management, and optimize resource utilization \cite{3gpp.37.817,3gpp.38.843,3gpp.38.744}. It is worth noting that 3GPP RAN has been developing its AI/ML framework that depicts the general aspects of an AI/ML workflow, and a 3GPP-wide alignment on the AI/ML framework as well as terminology has been underway \cite{3gpp.22.850}. It is expected that a unified AI/ML framework will be in place soon for the end-to-end system and across SDOs. 

O-RAN pioneered the architectural innovations towards an AI-driven RAN since its birth in 2018, highlighting an open and modular methodology. Some new components dedicated to RAN intelligence are introduced, namely the near-real-time RAN intelligent controller (Near-RT RIC) and the non-real-time RAN intelligent controller (Non-RT RIC) 
\cite{O-RAN.WG1.OAD,O-RAN.WG2.RICARCH,O-RAN.WG3.RICARCH}. Thanks to the open interfaces, O-RAN features fine-grained data collection in RAN, and extensive support for AI/ML workflows across Near-RT RIC and Non-RT RIC. In addition, both RICs feature modularity with their cloud-native software (termed xApp and rApp) that can be developed and deployed on demand, enabling light-weighted and customizable AI/ML solutions for RAN. With its RIC-enabled centralized optimization, O-RAN complements the 3GPP RAN with more AI/ML use cases \cite{O-RAN.WG3.UCR} including traffic steering, quality of service (QoS)/quality of experience (QoE) optimization, massive MIMO optimization, etc. 

Apart from 3GPP and O-RAN, several projects in ITU and ETSI are also relevant to the convergence of AI and RAN. ITU leads the global initiative on the overall AI/ML related frameworks for the telecom networks, providing guidance and methodology for various SDOs. Under the ITU umbrella, the ML5G Focus Group (FG-ML5G) was formed to specify the architectural framework for the application of AI/ML in future networks \cite{ITU.Y.3172}, and followed by the Autonomous Networks Focus Group (FG-AN) that aimed to minimize human interventions in network management and control from with the help of AI/ML models \cite{ITU.Y.3061}. In ETSI, Experiential Networked Intelligence (ENI) \cite{ENI} and Zero-Touch Service and Network Management (ZSM) \cite{ZSM} projects were conducted, where AI/ML is denoted as the key enabler for cross-domain and automated network management.

Thanks to the extensive standardization work in the past few years, people now have a better understanding of what are, and what are not, in the scope of standardization with respect to AI/ML. An AI/ML model is essentially an algorithm, but it is different from the non-AI/ML-based algorithms in that it can be generated (i.e., trained) from the data from RAN and UE more dynamically. With that in mind, most of the efforts in the standards so far are devoted to the management/control of AI/ML workflow, including but not limited to, model identification and meta, awareness of relevant RAN/UE capability, data collection aspects (protocols, data types and formats, etc.), mechanisms for AI/ML model delivery/transfer, deployment/update, activation/deactivation, and fallback to non-AI/ML-based algorithms. Besides, the additional requirements from dedicated AI/ML infrastructure may have an impact the RAN architecture, in order to optimize resource utilization and energy consumption for the MNOs. 
On the other hand, similar to the non-AI/ML-based algorithms, the weights of an AI/ML model and how it is trained are out of standardization scope. It remains for further investigation whether certain generic model structures (e.g., convolutional neural network, CNN) and associated complexity metrics can be standardized, to facilitate use cases where two-sided/shared models are generated and employed, though many vendors have concern on the disclosure of proprietary information.  

Among the remarkable achievements, a number of deficiencies in the 5G/5G-A integration of AI/ML can be observed from the following perspectives:

\begin{itemize}
    \item Organization of work: the standardization of AI/ML capabilities in the different WGs of 3GPP as well as other SDOs was launched with a bottom-up pattern, resulting in huge overhead to avoid inconsistencies in the end-to-end system. The coordination among the SDOs are also usually slow and inefficient, which may cause overlapping scope of work.  

    \item Architecture: the integration of AI/ML for 5G systems has been constrained by the existing architectural designs, which is basically an ``add-on" approach that limits any new functions/functionalities in making the best of AI/ML techniques.

    \item Use cases: The exploration of AI/ML-enabled use cases was started in the management and the core network domains and extended to RAN with selected use cases recently. More use cases remain to be investigated in the 6G era to fully unleash the potential of AI/ML in the link-level and area-level RAN performance. 

    \item AI/ML data collection: In front of the massive and distinct data collection requirements from various use cases and tasks (e.g., AI/ML model training and inference), the efficiency of data collection mechanisms counts in the overall system performance. This issue is especially severe in RAN, where both the existing control plane and user plane solutions have some limitations. Besides, controllability and observability in the process have been the concern of MNOs. Other aspects associated with AI/ML data, including cross-domain data collection, data cleaning/pre-processing and storage, also call for unified designs in the standards.

    \item AI/ML model: First, the 5G AI/ML capabilities are mostly based on off-line trained models, whose performance may deteriorate in a realistic environment when the distributions of training data and real-time input data are mismatched. To address that issue, online model training and fine-tuning aided by pre-validation is most promising. Second, the existing performance monitoring designs are limited to a per-use-case basis, and a more systematic approach remains to be established. Third, the application of AI/ML models in RAN is still in an early stage, with limited support, if any, for the cutting-edge model structures and paradigms like federated learning and large models. 

    \item AI computing resource: The impact of AI computing resource in the distributed scenarios should be emphasized in the standardization rather than left to implementations. The major MNOs are in deep concern with the utilization efficiency of AI computing resources in the distributed RAN nodes, as such hardware is typically expensive and energy-hungry.

    \item RAN services: While RAN is traditionally regarded as a dump pipe for communication services only, the emergence of AI applications and devices (e.g., AI wearables) that require ultra-low latency computing has opened new opportunities for 6G RAN. The 6G RAN can not only enhance user experience with awareness of the AI/ML service traffic, but also be leveraged to produce new types of services beyond communication, e.g, edge AI computing, which could create new sources of revenue for the MNOs. 
    
\end{itemize}

\begin{figure*} [ht]
\centering
	\includegraphics[scale=0.6]{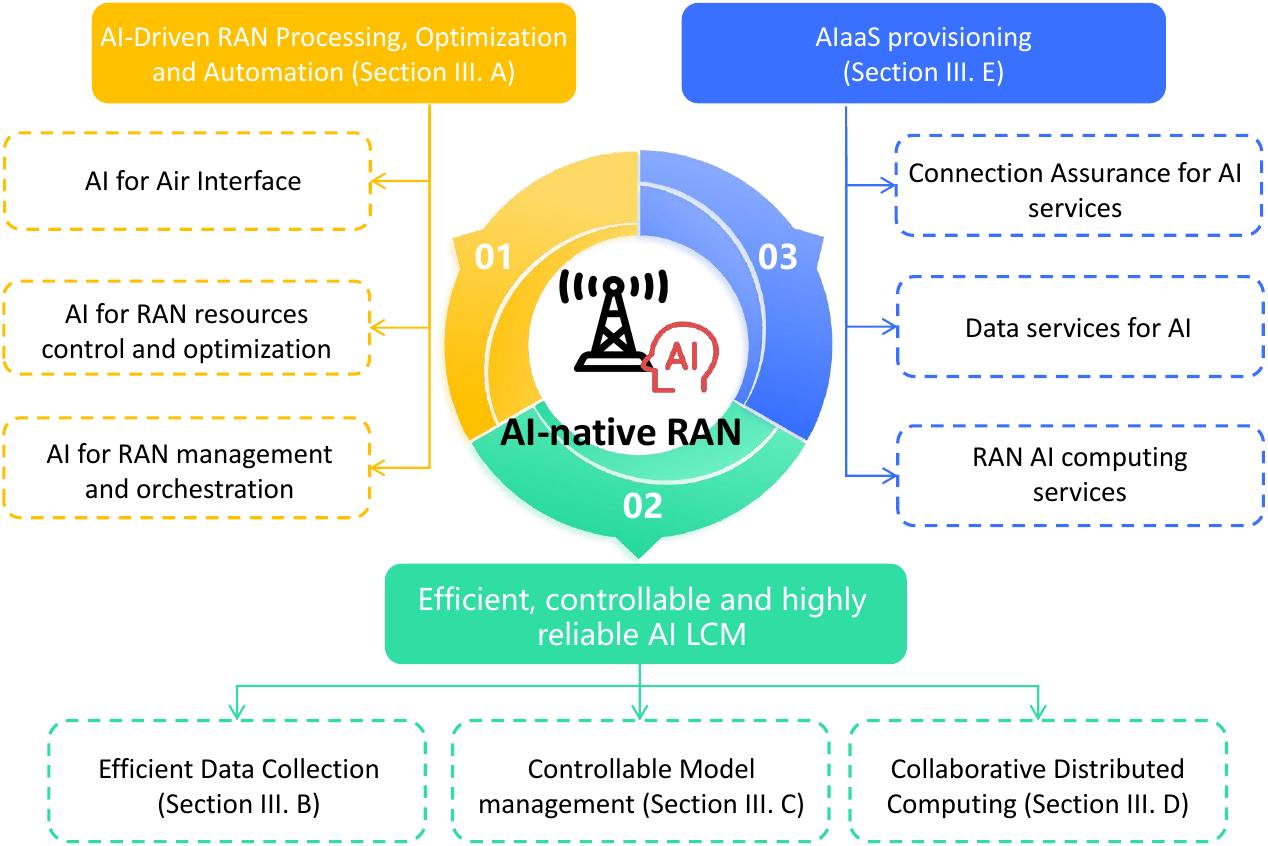}
\caption{Three essential capabilities of AI-Native RAN.}

\end{figure*}

\section{AI-Native RAN: Concept and Essential Capabilities}

\begin{table*}[]
\caption{Typical use cases for AI-driven RAN processing, optimization and automation}
\label{Table of use cases}
\begin{tabular}{|c|c|c|c|c|c|}
\hline
scenarios                                                                                                    & use case                                                                                                          & \begin{tabular}[c]{@{}c@{}}key challenges of\\ non AI-based solutions\end{tabular}                                                                                                                                                                              & \begin{tabular}[c]{@{}c@{}}AI inference\\ location\end{tabular}             & \begin{tabular}[c]{@{}c@{}}potential AI methods\\ \& parameter scale\end{tabular}                                                            & \begin{tabular}[c]{@{}c@{}}research \\ status\end{tabular}                                                          \\ \hline
\multirow{7}{*}{\begin{tabular}[c]{@{}c@{}}AI for RAN \\ resources control \\ and optimization\end{tabular}} & \begin{tabular}[c]{@{}c@{}}load \\ balancing\end{tabular}                                                         &                                                                                                                                                                                                                                                                 & BS-side                                                                     & \begin{tabular}[c]{@{}c@{}}DNN(\textless{}1M)\\ LSTM(\textless{}10M)\\ DRL(\textless{}10M)\end{tabular}                                      & \begin{tabular}[c]{@{}c@{}}5G-A 3GPP Release 18,\\ O-RAN\end{tabular}                                               \\ \cline{2-2} \cline{4-6} 
                                                                                                             & \begin{tabular}[c]{@{}c@{}}mobility \\ management\end{tabular}                                                    & \multirow{6}{*}{\begin{tabular}[c]{@{}c@{}}passive optimization \\ degrades the \\ performance;\\ high-dimensional, \\ complex \\ and often non-convex \\ optimization problems; \\ hard to find the \\ optimal solution \\ in a real time manner\end{tabular}} & \begin{tabular}[c]{@{}c@{}}UE-side,\\ BS-side\end{tabular}                  & \begin{tabular}[c]{@{}c@{}}random forest \\ gradient boosting \\ LSTM(\textless{}1M)\end{tabular}                                            & \begin{tabular}[c]{@{}c@{}}5G-A 3GPP Release 18/19,\\ O-RAN\end{tabular}                                            \\ \cline{2-2} \cline{4-6} 
                                                                                                             & \begin{tabular}[c]{@{}c@{}}network \\ energy saving\end{tabular}                                                  &                                                                                                                                                                                                                                                                 & \multirow{5}{*}{BS-side}                                                    & \begin{tabular}[c]{@{}c@{}}DNN(\textless{}1M)\\ LSTM(\textless{}10M)\\ DRL(\textless{}10M)\end{tabular}                                      & \begin{tabular}[c]{@{}c@{}}5G-A 3GPP Release 18,\\ field trial in 5G-A \\ commercial network,\\ O-RAN\end{tabular}  \\ \cline{2-2} \cline{5-6} 
                                                                                                             & network slicing                                                                                                   &                                                                                                                                                                                                                                                                 &                                                                             & \begin{tabular}[c]{@{}c@{}}LSTM(\textless{}10M)\\ GRU(\textless{}10M)\\ Transformer(10M-1B)\\ double dueling DQN(\textless{}1M)\end{tabular} & \multirow{2}{*}{\begin{tabular}[c]{@{}c@{}}5G-A 3GPP Release 19,\\ O-RAN\end{tabular}}                              \\ \cline{2-2} \cline{5-5}
                                                                                                             & \begin{tabular}[c]{@{}c@{}}capacity \\ and coverage \\ optimization\end{tabular}                                  &                                                                                                                                                                                                                                                                 &                                                                             & \begin{tabular}[c]{@{}c@{}}DRL(\textless{}10M)\\ GNN(\textless{}10M) \\ LSTM(\textless{}10M)\\ Transformer(10M-1B)\end{tabular}              &                                                                                                                     \\ \cline{2-2} \cline{5-6} 
                                                                                                             & \begin{tabular}[c]{@{}c@{}}multi-user \\ scheduling and \\ resources allocation\end{tabular}                      &                                                                                                                                                                                                                                                                 &                                                                             & \begin{tabular}[c]{@{}c@{}}DNN(\textless{}1M)\\ DRL(\textless{}10M)\end{tabular}                                                             & \begin{tabular}[c]{@{}c@{}}6G candidate use case,\\ O-RAN\end{tabular}                                              \\ \cline{2-2} \cline{5-6} 
                                                                                                             & \begin{tabular}[c]{@{}c@{}}QoS/QoE optimization \\ and service level \\ agreement \\ (SLA) assurance\end{tabular} &                                                                                                                                                                                                                                                                 &                                                                             & \begin{tabular}[c]{@{}c@{}}k-means(\textless{}1M)\\ DNN(\textless{}1M)\\ CNN(\textless{}1M)\\ DRL(\textless{}10M)\end{tabular}               & \begin{tabular}[c]{@{}c@{}}field trial in 5G-A \\ commercial network,\\ 6G candidate use case,\\ O-RAN\end{tabular} \\ \hline
\multirow{8}{*}{\begin{tabular}[c]{@{}c@{}}AI for \\ air interface\end{tabular}}                             & positioning                                                                                                       & limited accuracy                                                                                                                                                                                                                                                & \begin{tabular}[c]{@{}c@{}}UE-side,\\ BS-side, \\ LMF-side\end{tabular}     & \begin{tabular}[c]{@{}c@{}}DNN(\textless{}30M) \\ CNN(\textless{}30M)\end{tabular}                                                           & \multirow{4}{*}{5G-A 3GPP Release 18/19}                                                                            \\ \cline{2-5}
                                                                                                             & beam management                                                                                                   & \begin{tabular}[c]{@{}c@{}}limited accuracy in \\ medium and \\ high-speed scenario; \\ unacceptable overhead \\ for reasonable accuracy\end{tabular}                                                                                                           & \begin{tabular}[c]{@{}c@{}}BS-side, \\ UE-side\end{tabular}                 & \begin{tabular}[c]{@{}c@{}}CNN(\textless{}10M)\\ LSTM(\textless{}10M)\\ Transformers(\textless{}10M)\end{tabular}                            &                                                                                                                     \\ \cline{2-5}
                                                                                                             & CSI compression                                                                                                   & \begin{tabular}[c]{@{}c@{}}theoretically intractable\\ to model/formulate\end{tabular}                                                                                                                                                                          & \begin{tabular}[c]{@{}c@{}}two-sided \\ model\end{tabular}                  & \begin{tabular}[c]{@{}c@{}}CNN(\textless{}10M)\\ LSTM(\textless{}10M)\\ Transformers(\textless{}10M)\end{tabular}                            &                                                                                                                     \\ \cline{2-5}
                                                                                                             & CSI prediction                                                                                                    & \begin{tabular}[c]{@{}c@{}}outdated and \\ limited accuracy\end{tabular}                                                                                                                                                                                        & \multirow{4}{*}{\begin{tabular}[c]{@{}c@{}}BS-Side,\\ UE-side\end{tabular}} & \begin{tabular}[c]{@{}c@{}}MLP-Mixer(\textless{}10M)\\ 2D-FCN(\textless{}1M)\\ CNN(\textless{}1M)\end{tabular}                               &                                                                                                                     \\ \cline{2-3} \cline{5-6} 
                                                                                                             & \begin{tabular}[c]{@{}c@{}}AI based \\ receiver /\\ transceiver\end{tabular}                                      & \begin{tabular}[c]{@{}c@{}}theoretically intractable\\ to model/formulate\end{tabular}                                                                                                                                                                          &                                                                             & \begin{tabular}[c]{@{}c@{}}ESN(\textless{}10M)\\ Transformers(\textless{}10M)\end{tabular}                                                   & \multirow{4}{*}{6G candidate use case}                                                                              \\ \cline{2-3} \cline{5-5}
                                                                                                             & \begin{tabular}[c]{@{}c@{}}PA non-linearity \\ compensation\end{tabular}                                          & \begin{tabular}[c]{@{}c@{}}theoretically intractable\\ to model/formulate\end{tabular}                                                                                                                                                                          &                                                                             & ESN(\textless{}10M)                                                                                                                          &                                                                                                                     \\ \cline{2-3} \cline{5-5}
                                                                                                             & \begin{tabular}[c]{@{}c@{}}Interference \\ prediction \\ and handling\end{tabular}                                & \begin{tabular}[c]{@{}c@{}}outdated and \\ limited accuracy\end{tabular}                                                                                                                                                                                        &                                                                             & \begin{tabular}[c]{@{}c@{}}LSTM(\textless{}10M)\\ Transformers(\textless{}10M)\end{tabular}                                                  &                                                                                                                     \\ \cline{2-5}
                                                                                                             & \begin{tabular}[c]{@{}c@{}}RS overhead \\ reduction \\ (incl. superimposed \\ pilot)\end{tabular}                 & \begin{tabular}[c]{@{}c@{}}high reference \\ signal overhead\end{tabular}                                                                                                                                                                                       & BS-Side                                                                     & \begin{tabular}[c]{@{}c@{}}CNN(\textless{}10M)\\ LSTM(\textless{}10M)\\ Transformers(\textless{}10M)\end{tabular}                            &                                                                                                                     \\ \hline
\end{tabular}
\end{table*}

The experience gained from AI/ML integration in 5G and 5G-A highlights both the potential and the limitations of current approaches. While notable progress has been made, the fragmented and case-specific nature of existing work reveals the absence of a unified, long-term architectural vision for the deep fusion of AI and RAN. This calls for a fundamental paradigm shift in 6G design, where AI is not merely an add-on, but a native capability embedded throughout the RAN. A well-articulated concept of AI-Native RAN is essential to align standardization efforts, steer architectural evolution, and provide clear, unified guidance for future product development. This mirrors the pivotal role played by the Cloud Native Computing Foundation (CNCF) in shaping the standardized concept of cloud-native computing, which subsequently laid the foundation for cloud evolution.

From an operator’s perspective, we attempt to articulate a concept of AI-Native RAN, inspired by the cloud-native paradigm. In this view, AI-Native RAN is a radio access network fundamentally driven by AI, with AI deeply embedded in its functions, architectures, protocols, and operations. Such integration empowers intelligent RAN optimization and automation, enabling real-time, efficient, and diversified service delivery across communication, computing, and AI tasks for ubiquitous applications.

As illustrated in Fig. 1, realizing the vision of AI-Native RAN requires three essential capabilities. The first is AI-driven RAN processing, optimization, and automation, which enables intelligent control across the air interface, RAN management and resource optimization; this will be discussed in Section III.A. The second is efficient, controllable, and reliable AI lifecycle management (LCM), covering data collection, model management, and collaborative computing, which will be detailed in Sections III.B, III.C, and III.D, respectively. The third is AI-as-a-Service (AIaaS) provisioning, which transforms the RAN into a unified platform for both communication and AI applications, and will be introduced in Section III.E.

\subsection{AI-Driven RAN Processing, Optimization, and Automation}

With the increasing complexity of modern wireless networks, traditional rule-based or model-driven RAN optimization methods are facing growing challenges. In contrast, AI technologies demonstrate unique advantages in handling complexity, scale, and uncertainty. In the context of RAN processing, optimization, and automation, AI offers three key capabilities: the ability to learn in scenarios where explicit modeling is infeasible, the capability to make real-time decisions in large-scale, high-dimensional environments, and the power to perform accurate prediction and proactive adaptation based on complex spatiotemporal patterns for proactive optimization.

Overall, these capabilities position AI as a powerful enabler across physical layer (L1), data link layer (L2) and network layer (L3).  
Table \ref{Table of use cases} shows typical use cases of AI for air interface (L1), and AI for resources control and optimization (L2/L3) in a systematic view.

\begin{figure}
\centering
	\includegraphics[scale=1.3]{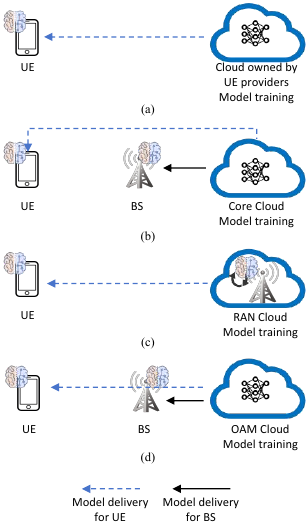}
\caption{Four potential AI model training, delivery and deploy modes in AI-Native RAN. (a) UE cloud training; (b) core network training; (c) BS training; (d) operation administration and maintenance (OAM) cloud training.}
\label{4int}
\end{figure}

The models applied to these use cases must first undergo training, testing, and validation to ensure the performance of the aforementioned AI-driven RAN processing, optimization, and automation in practical systems. 
Fig. \ref{4int} shows the four typical approaches on how AI model can be trained, delivered and deployed in RAN.

\textbf{Standardization Considerations:} 
5G standardization has pioneered valuable AI applications in single RAN modules, laying the groundwork for broader AI integration. In 6G, it is expected to advance beyond single-module toward more complex and high-value use cases, such as joint optimization across multiple modules and cross-layer AI coordination, aiming to fully harness AI for ultra-reliable and high-performance user experiences. Promising use cases may include AI-based transceivers, power amplifier (PA) nonlinearity compensation and reference signal (RS) overhead reduction, etc. To support the prioritization of use cases, a comprehensive and systematic performance evaluation framework is essential. It should go beyond basic accuracy or throughput metrics. Instead, it must incorporate multi-dimensional criteria, including energy efficiency, model generalization capability, and overall system cost, which encompasses AI-specific overheads such as computation, data collection, and storage. While specific models and algorithms will typically remain outside the scope of standardization, key enablers—such as data collection mechanisms, model management procedures, and distributed AI computing frameworks—need to be standardized for scalable and interoperable AI-native RAN deployments.

\subsection{Efficient Data collection}
\label{data collection mechanisms}
The current 5G RAN protocol design is oriented to the communication purposes but less to the needs of AI/ML data collection. A comprehensive review of existing data collection approaches was conducted in 3GPP RAN \cite{3gpp.38.843}.  
On the air interface, the control plane suffers from scalability in the maximum payload size. While the user plane has fewer restrictions in the traffic data volume, the data flows are terminated by the user plane function (UPF) in the core network and may be subject to charging, making it less favorable for AI/ML tasks in RAN. Besides, when a large volume of data is collected over the air, there exists a risk in draining the radio link and impeding the basic connectivity services in RAN. In addition, multiple aspects, including data privacy and security, also pose challenges for data collection. In Release 19, the different paths for data collection from UE (data collection server in OTT, in the core or in the OAM) are under discussion in 3GPP RAN under the existing 5G framework \cite{QC3GPPAI}. However, when it evolves to 6G, new designs shall be considered for native data collection support in RAN. 

\subsubsection{\textbf{Dedicated AI radio bearer}} 
\begin{figure} 
\centering
	\includegraphics[width=0.45\textwidth]{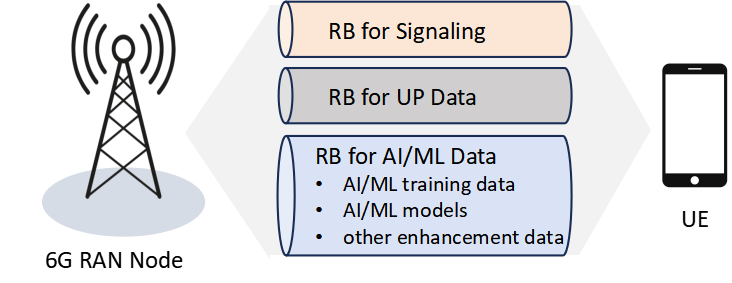}
\caption{AI radio bearer.}
\label{AIB}

\end{figure}

A novel type of radio bearers \cite{CMCCnewRB} may be defined to support AI/ML related data delivery on the air interface as shown in Fig. \ref{AIB}. 
Such a new radio bearer is solely managed by RAN, which allows RAN to establish and configure it dynamically upon the demand of AI/ML use cases in 6G RAN. 
As per the payload, the radio bearer is capable of multiple AI/ML related data types, including but not limited to, the AI/ML training data, the AI/ML models, and other enhancement data that may assist the AI/ML workflows (e.g., the performance metrics of AI/ML models). 
It is expected that the new radio bearers will ensure both flexibility and efficiency in the AI/ML workflows in 6G RAN. 
In addition to the new radio bearer, a new data tunnel may be specified on the network interfaces that interconnect the 6G RAN nodes, to enable the exchange of AI/ML related data. For example, a 6G RAN node can collect data from adjacent nodes or their served UEs for centralized model training, and thereafter, the node can deliver the trained models to adjacent nodes or the UEs on demand. 

\subsubsection{\textbf{Cross-domain data collection}} 
\begin{figure} [h]
\centering
	\includegraphics[width=0.45\textwidth, height=3.5cm]{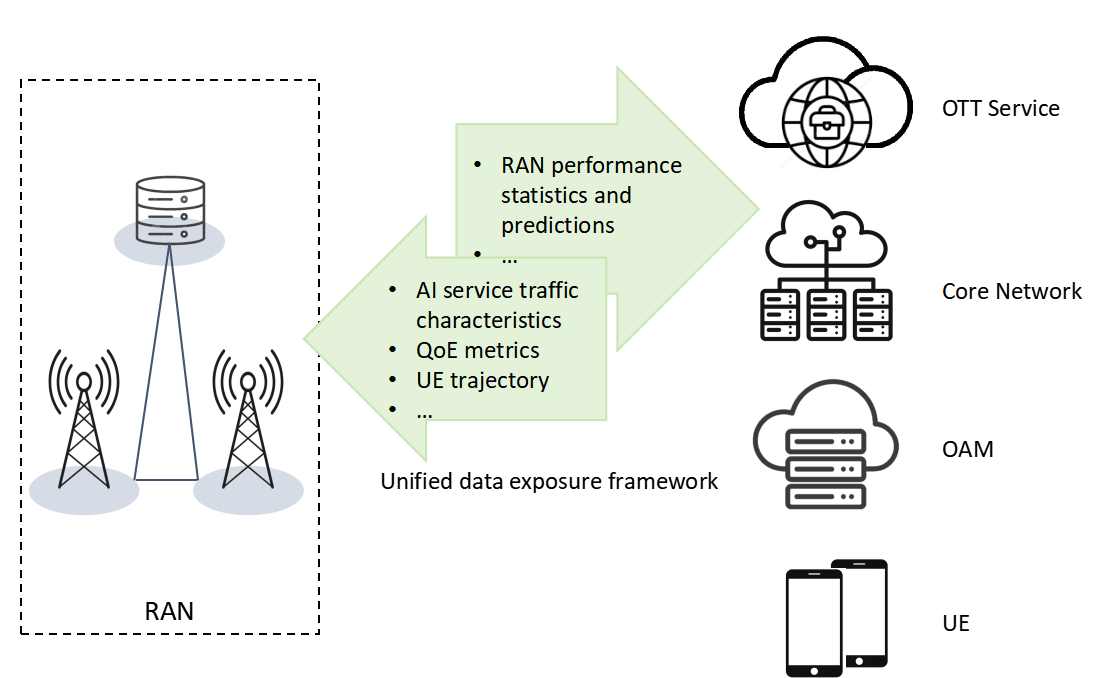}
\caption{Cross-domain data collection.}
\label{CD}
\end{figure}

The 6G AI-Native RAN is expected to enable cross-domain AI/ML for end-to-end service assurance, which will entail cross-domain data collection. One use case is the service-aware RAN optimization for AI services. 
The 6G RAN can optimize its behavior by acquiring the real-time traffic characteristics from the application servers. The QoE metrics for AI/ML service may also be useful in assisting RAN to improve the user’s experience in a more specialized approach, for instance, the time to first token (TTFT) in large language model (LLM) inference.
Upon the service-layer information, the AI-Native RAN is able to employ service-specific optimizations such as PDU set mapping and DRB configuration. Meanwhile, the real-time RAN performance metrics and predictions may also be collected by the core network to adjust its QoS configurations and UPF behavior more dynamically. With the diverse data sources and collectors, it would be more efficient for 6G RAN to provide a holistic framework for cross-domain data collection, which is beyond the 5G network capability exposure. Fig. \ref{CD} illustrates the cross-domain data collection envisioned in 6G. 

\subsubsection{\textbf{Task-driven customizable RAN data collection mechanism}} 
\begin{figure} [h]
\centering
	\includegraphics[width=0.45\textwidth]{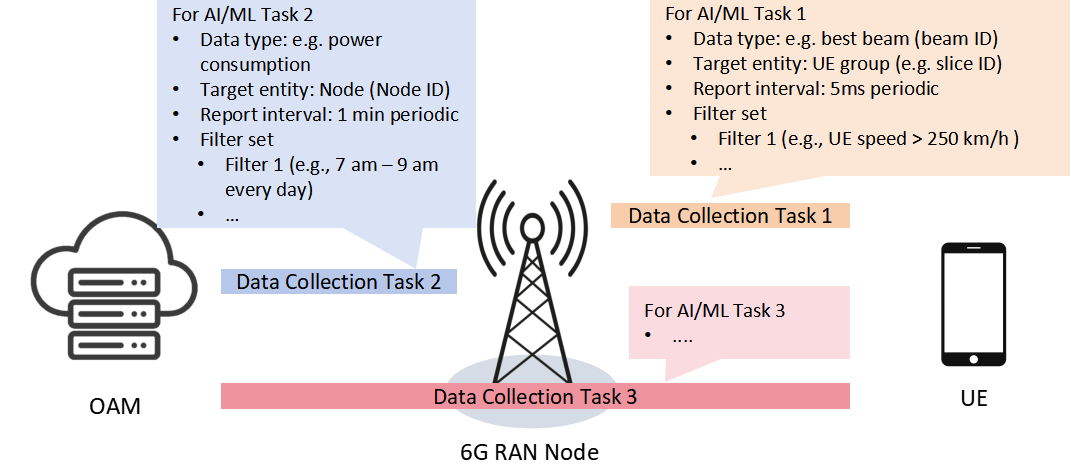}
\caption{Task-driven customizable RAN data collection mechanism.}
\label{FDC}
\end{figure}

Beyond minimization of drive test (MDT), performance measurements (PMs) and measurement reports (MRs), a new data collection mechanism is essential to fulfill the diverse requirements on data collection in 6G RAN. The new mechanism highlights real-time, fine-granularity and flexibility. Compared with the OAM-oriented mechanisms that work with about $15$ minutes report interval, the new mechanism controlled by the 6G RAN nodes can support a latency of 10 ms or lower, for the real-time AI/ML inference tasks. It is also capable of RAN data collection on a specific UE, so as to enable more fine-grained AI/ML optimization tasks. Moreover, the new mechanism is specialized to the reduction of unnecessary data collection, e.g., to train an AI/ML model for high-speed UEs, it would be a waste of radio resources to collect UE information from all the users in a cell. For that purpose, the new mechanism could support a customizable set of filters to ensure only the needed RAN data are collected. Apart from the data type, a filter may specify the associated service type (e.g., throughput with a specific 5QI), area of interest, time range (e.g., day-time) or UE status (e.g., not to collect data from low-battery UEs). Such a filter may also specify a criterion consisting of math symbols (e.g., greater than “$>$”) and threshold values, e.g., “when UE velocity $>$ 250 km/h”. The logical expression is evaluated at the data source, with which only the subset of the RAN data that satisfies the criterion is delivered to the data collector.
Fig. \ref{FDC} illustrates the typical examples of a task-driven customizable RAN data collection mechanism. 

\textbf{Standardization Consideration:} The impact of data collection on 6G RAN standardization is twofold. One aspect is the generic protocols on relevant interfaces including the air interface (i.e., Uu), the RAN interfaces (e.g., Xn/F1/E1 interfaces in 5G), and the interfaces with OAM and core network. The design of such protocols shall not only ensure efficiency for diverse data collection tasks, but also MNO's observability and controllability. The other aspect is the detailed specifications of AI/ML related data types and formats (including the input/output data for model inference, training data, etc.), and the associated configurations for the data collection tasks. Such specifications will be consistently built on top of the aforementioned generic protocols and driven by specific scenarios and use cases. 
Beyond data collection, aspects including data pre-processing, storage and retrieval need more attention in the standards, which may require a more comprehensive data management framework \cite{DataPlane}.

\subsection{Controllable and Highly Reliable AI Model Management}

\begin{figure*} 
\centering
	\includegraphics[width=0.9\textwidth, height=9cm]{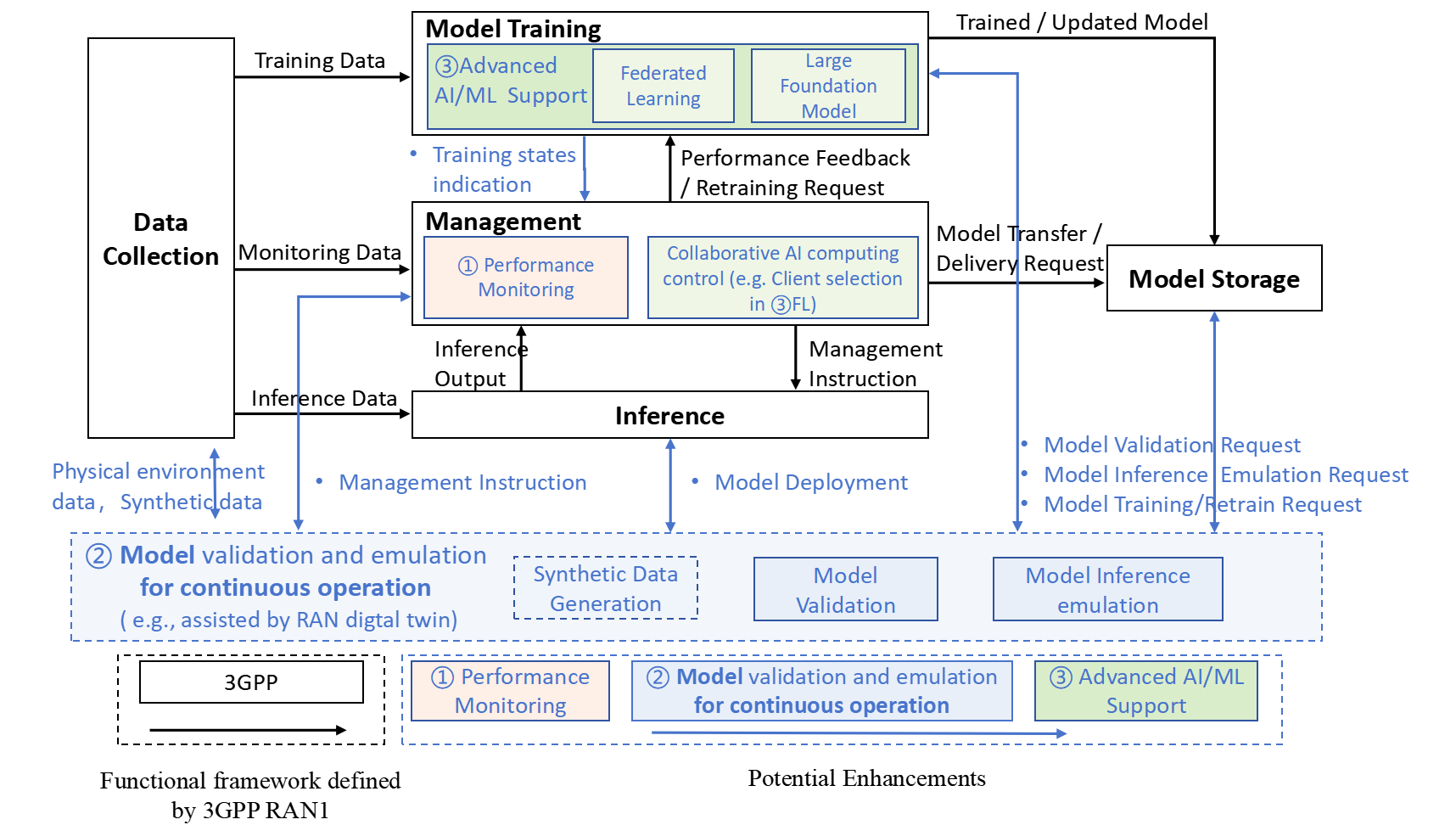}
\caption{Enhanced AI/ML model management for 6G RAN.}
\label{management}
\end{figure*}

As AI becomes integral to 6G RAN control and optimization, robust and controllable AI model management is critical due to the probabilistic nature of AI, which complicates predictability in dynamic environments. While 3GPP has laid the groundwork for AI/ML management in 5G, covering data collection, training, inference, and monitoring, current frameworks fall short for commercial deployment, lacking operational drift detection, pre-deployment validation, and model continuous learning to adapt to changing conditions. In addition, emerging techniques like federated learning and large foundation models offer improved accuracy and adaptability. To enable AI-Native 6G RAN, advancements for model management are expected. Fig.~\ref{management} illustrates an enhanced functional model management framework in 6G, building upon the 5G AI/ML framework.

\textbf{Enhanced Model Performance Monitoring}: Once deployed in the RAN, AI models must be continuously monitored and tuned to ensure their stability and reliability. While 3GPP has initiated preliminary studies on AI model performance monitoring in 5G/5G-A, focusing on inference accuracy, system KPIs, and data distribution \cite{3gpp.38.843}, AI-Native RAN requires a three-dimensional framework encompassing AI model performance, network performance, and resource utilization. Silent degradation from data drift, unanticipated scenarios, or traffic variations necessitates end-to-end monitoring that integrates model accuracy metrics with network KPIs (e.g., latency, throughput) to assess real-world impact. Resource indicators (inference delay, memory consumption, energy efficiency) are also critical for operational feasibility and cost evaluation, forming a foundation for standardized monitoring frameworks. Representative indicators are illustrated in Table \ref{performance}, which provides a foundation for standardized and scalable monitoring frameworks. These metrics are expected to be collected from diverse AI inference nodes and interfaces across RAN.

\begin{table*}
\centering
\caption{Examples of the ML performance monitoring metrics.}
\label{performance}
\begin{tabular*}{0.76\linewidth}{@{}ll@{}}
Dimension&Example Metrics\\\hline
\multirow{4}{*}{Model Performance}&Accuracy, precision, recall, MAE, MSE, RMSE, etc.  \\
~&Generalization ability, e.g., Adversarial attack success rate\\
~&Calculation efficiency, e.g., Inference latency\\
~&Output error distribution/residual distribution\\\hline
\multirow{3}{*}{Network Performance}&Network energy saving: Energy consumption/efficiency, packet loss rate, user data rate etc.\\
~&Mobility management: Handover failure, radio link failure, delay, packet loss rate etc.\\
~&CSI compression: BLER, average user throughput (UPT), 5\%-UPT etc.\\\hline
\multirow{3}{*}{Resource Performance}&Computational load, e.g. CPU/GPU utilization \\
~&Storage overhead, e.g., Data storage/cache occupancy\\
~&UE and RAN energy usage status\\\hline
\end{tabular*}
\end{table*}

\textbf{Model Continuous Operation}: 
Model continuous operation is essential for ensuring robust and adaptive AI behavior in AI-native RAN, particularly under dynamic network conditions and user mobility. As UEs move across cells, consistent inference must be maintained through synchronization of model context (e.g., model ID, version) across gNBs, supported by standardized Xn or OAM-gNB interfaces. To adapt to changing environments, the network may adopt performance-triggered retraining, where model degradation detected by the gNB or UE initiates data collection, retraining, validation, and inference emulation before redeployment, or periodic retraining, which proactively updates models at regular intervals using refreshed datasets. Both approaches require meticulous model version control and dataset traceability to ensure consistency between training and inference. To further enhance robustness, RAN Digital Twins (DTs) can provide synergetic data and support model validation by simulating real-world scenarios, enabling pre-deployment testing, synthetic data generation, and closed-loop optimization. This integrated framework supports safer, more reliable AI deployment and sustainable model evolution in future 6G networks.

\textbf{Advanced AI/ML Technology Support:} 
Federated learning (FL) enhances AI model generalization in 6G RAN by enabling privacy-preserving collaborative training across distributed gNBs. Each node trains a local model on its own data, while updates are aggregated—either through peer coordination or centralized RAN nodes—to form a global model that reflects diverse network conditions, such as varying traffic loads and mobility patterns. Effective deployment of FL in RAN requires architectural support for adaptive participant selection to prioritize nodes with high-value data, mechanisms to handle device and data heterogeneity for stable and timely convergence, and efficient update aggregation. These considerations are essential to realize scalable and robust FL workflows in AI-native RAN environments. Large foundation models offer a unified approach to modeling complex radio dynamics in 6G RAN, enabling tasks like channel prediction and beam selection with strong generalization across diversified environments. To enable it, efficient distributed adaptation, model compression for low-latency inference, and task-aware monitoring need to be considered so as to ensure performance in resource-limited settings.

\textbf{Standardization Consideration:} To ensure the stability and reliability of AI models in 6G RAN, it is essential to define the collection of multi-dimensional performance monitoring metrics through interfaces such as Uu, Xn, and those associated with network management. Based on the monitored performance, standardized procedures and interfaces are required to enable the exchange of model management instructions, such as triggering model retraining, updating model versions, or reverting to non-AI baselines, among UEs, RAN nodes, and OAM systems. Building on this foundation, end-to-end model lifecycle management must be supported through interoperable and standardized workflows, covering retraining, validation, and deployment. In addition, mechanisms for model and dataset ID traceability, version control, and lifecycle metadata management should be specified to ensure consistent and reliable AI behavior across multi-vendor deployments. To future-proof the architecture, an extensible framework is also needed to accommodate emerging AI/ML technologies, such as federated learning (FL), and to support their seamless integration into the 6G RAN ecosystem.

\begin{figure*} 
\centering
	\includegraphics[width=0.9\textwidth, height=4.5cm]{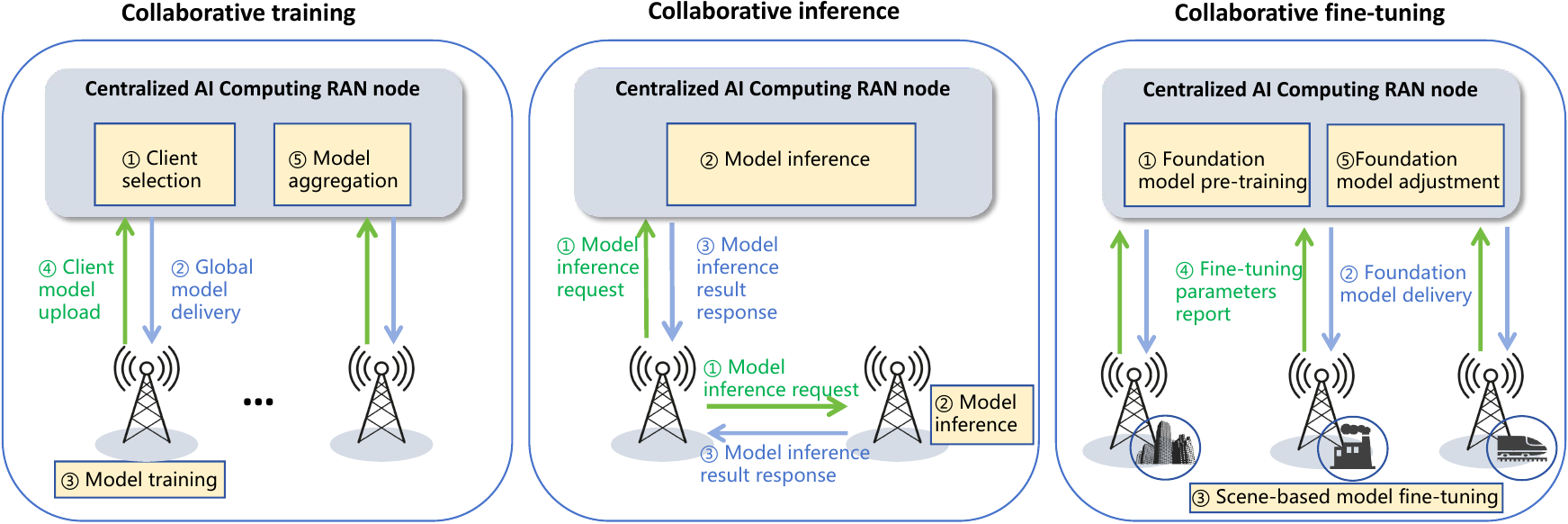}
\caption{Three types of collaborative AI computing in AI-Native RAN.}
\label{collaborative}
\end{figure*}

\subsection{Distributed Collaborative AI computing}
The integration of AI into 6G RAN promises to revolutionize network optimization and service delivery. However, widespread deployment is limited by constrained computational resources, energy budgets, and fragmented, localized data at individual nodes. These factors hinder the training and execution of robust AI models, underscoring the need for a distributed, collaborative computing paradigm across the RAN. To overcome these challenges, centralized AI computing RAN nodes and coordination functions can be introduced to enable AI resource pooling and orchestrate workloads across multiple gNBs and UEs. Effective collaborative AI mechanisms must be developed to optimize resource utilization and unify intelligence across heterogeneous network elements, including gNBs, UEs, and centralized AI nodes. Three fundamental collaborative modes are then envisioned: collaborative training, inference, and fine-tuning, as shown in Fig.~\ref{collaborative}. 

\textbf{Collaborative Training:}
gNBs or a dedicated centralized AI computing RAN node can collaboratively train global or personalized AI models using distributed datasets, typically via federated or split learning paradigms. This approach mitigates data sparsity and heterogeneity at individual nodes while preserving user data privacy and reducing communication overhead. For instance, in a federated learning scenario for user mobility prediction, each gNB collects localized mobility data (e.g., handover logs, signal strength traces) and trains a local model on-site. The centralized AI computing RAN node aggregates these updates to refine a global mobility model, which is then redistributed to gNBs for subsequent training rounds. This iterative training loop ensures improved model generalization across the network.

\textbf{Collaborative Inference:}
Inference workloads can be flexibly partitioned between gNBs and centralized AI computing nodes to optimize performance under resource constraints. While capable gNBs and UEs can handle lightweight inference tasks locally, resource-limited gNBs can directly request inference from the centralized node. Leveraging aggregated data or intermediate inferences from multiple sites, the centralized AI computing node applies more advanced algorithms to generate globally optimized strategies, such as interference mitigation and dynamic resource allocation, which are then distributed to gNBs for coordinated execution. This collaborative paradigm enhances adaptability, efficiency, and overall network intelligence in real time.

\textbf{Collaborative Pre-Train and Fine-Tuning:} Pre-trained global models are fine-tuned at edge nodes using localized data, allowing adaptation to specific conditions while contributing updates to the shared global model. For instance, a central node distributes a general traffic classification model, which gNBs then fine-tune based on local traffic patterns to enhance scheduling and QoS enforcement. UEs also adapt QoE models with personalized feedback, enabling real-time optimizations like video bitrate adjustment. Fine-tuned updates are selectively sent back to improve the global model.

To support this distributed paradigm, mechanisms such as efficient resource pooling and dynamic scheduling are essential. gNBs and UEs can register compute capabilities with centralized RAN nodes, which form logical compute pools based on geography, hardware type, or service needs. These pools abstract heterogeneity and support the scalable deployment of AI workloads. Intelligent scheduling dynamically allocates resources like CPUs/GPUs based on task priority, QoS, and energy efficiency, ensuring optimal utilization across the RAN.

\textbf{Standardization Consideration}: 
Enabling collaborative AI computing in the RAN requires architectural definitions of distributed intelligence functions across gNBs and centralized AI computing nodes. This includes standardized procedures and interfaces to support collaborative training, inference, and model adaptation. A dedicated AI computing resource management and orchestration framework is also essential, covering functions like resource discovery, usage monitoring, and topology management. To ensure efficient coordination, interfaces between the management system and the RAN computing infrastructure must enable real-time computing resource reporting, AI workload scaling and migration, and seamless integration with existing RAN management systems across heterogeneous network nodes.

\subsection{AIaaS provisioning}

\begin{figure*} 
\centering
    \includegraphics[width=0.75\textwidth, height=6cm]{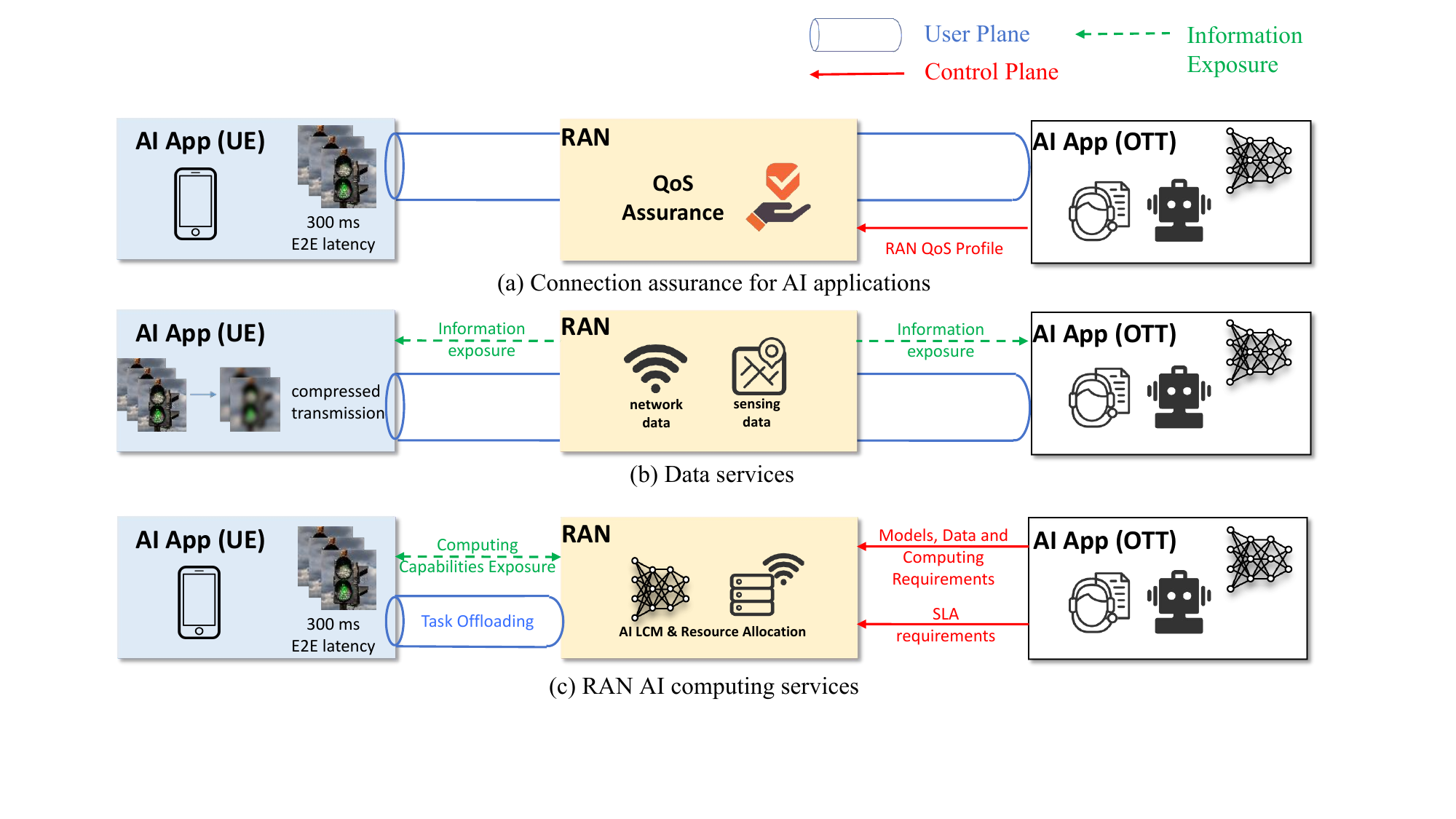}
\caption{Three AI application-oriented service modes enabled by AI-Native RAN. (a) connection assurance for AI applications; (b) data services; (c) RAN AI computing services.}
\label{3ser}

\end{figure*}

AIaaS is widely recognized as one of the key defining features of future 6G systems. Under this paradigm, network operators are expected to evolve from mere connectivity providers to edge AI computing providers, and ultimately to edge AI service enablers. Specifically, AI-Native RAN plays a pivotal role in facilitating this evolution, enabling three modes of service delivery that are specifically tailored to meet the needs of AI applications, depicted in Fig. \ref{3ser}.

\begin{itemize}
    \item Connection assurance for AI services refers to the RAN’s ability to deliver reliable, adaptive transmission tailored to the unique demands of AI workloads, requiring evolution beyond traditional network capabilities. Three key directions are essential: (1) enhancing the QoS framework to handle complex, opaque data flows like multimodal inputs and model updates; (2) supporting novel traffic patterns, such as lateral flows in embodied intelligence and uplink-heavy, bursty traffic in personal assistants; and (3) ensuring deterministic latency—e.g., under 200ms for assistants and below 100ms for embodied AI—potentially \cite{qu2025mobile} through selective preemption of background traffic.

\item Data services enable AI applications to adapt in real time for user experience by exposing internal network metrics (e.g., traffic load, link quality) and sensing data (e.g., mobility patterns, location). For example, personal assistant apps can adjust video compression rates for uplink transmission to reduce latency, while robotic systems may tune sensing and update frequencies to avoid packet loss. To support this, the RAN must also offer capabilities such as data prediction, augmentation, and cleansing to ensure timely and accurate data exposure. 

\item RAN AI computing services support external edge AI applications by exposing heterogeneous computing resources, such as GPUs and NPUs, within the RAN, while also meeting the internal demands of RAN intelligence. This dual use demands efficient scheduling, parallel processing, and tight coordination between communication and computing to ensure performance and user experience. For example, Ref.~\cite{ccjs} proposes a radio channel adaptive communication-computing joint optimization (RCACCJO) algorithm that dynamically selects model partition and early-exit points while allocating compute and uplink resources based on real-time channel states. Simulations show a 0.4\% accuracy gain over non-coordinated baselines like uniform bandwidth allocation.
\end{itemize}

\textbf{Standardization Consideration}:
To fully unlock the commercial potential of RAN’s connection capabilities, data assets, and infrastructure, it is essential to standardize QoS assurance mechanisms for AI, RAN data exposure, and the framework for provisioning low-latency AI computing services. Enabling RAN AIaaS requires an open API framework that facilitates seamless access to RAN resources. Through standardized APIs, operators can lower deployment costs, enhance operational efficiency, and deliver modular AI functions as on-demand services. Moreover, open APIs promote a collaborative ecosystem by enabling integration of third-party AI capabilities, thereby accelerating the evolution toward intelligent and flexible networks.

\section{AI-Native RAN Standardization for 6G}
In this section, the holistic vision for AI-Native RAN standardization in 6G will be presented with the emphasis on the Day 1 work, including the overall considerations, the key designs, and the collaboration between 3GPP and other SDOs. 

\subsection{Overall considerations for 6G AI-Native RAN}
\subsubsection{\textbf{The scope of 6G’s Day 1 standards will determine the efficacy of AI-Native RAN}}
The first release in each generation of 3GPP standards determines, to a tremendous extent, the practical capabilities of the commercial network of that generation. While the evolution of 3GPP standards typically takes 1.5 to 2 years per release, the operators upgrade their networks far less frequently due to the high total cost of ownership (TCO). In the 4G/5G era, the first release (i.e., Release 8 and Release 15, respectively) was the choice for large-scale deployment by the major operators, and accordingly gained the most R\&D support from the whole industrial chain. As a comparison, despite the numerous new features added in the subsequent releases in 4G/5G, only a few of them have been supported by the RAN and UE vendors, and even fewer purchased by operators. The same situation is likely to occur in 6G as well, leading to low utilization of the functionalities that arrive late.

One such example is the high-speed railway (HST) enhancement. China has an HST network which spans 48,000 kilometers as of 2024 and operates at a speed of up to 350 km/h. The radio signal in such environments suffers from a severe Doppler frequency shift. The passenger’s experience with the UE deteriorated significantly, due to the insufficient detection and compensation capabilities for Doppler frequency shift in the 4G and early 5G UEs as well as base stations. The related enhancements were introduced only in Release 16 and thereafter, making them less available in the live network. 

\subsubsection{\textbf{Forward compatibility is essential for the standardization of AI-Native RAN}} 
While backward compatibility is a must for virtually all the telecom standards, forward compatibility is of special significance for the standardization of AI-Native RAN. Such forward compatibility will have an impact on both hardware and software. 

From the hardware perspective, voices from many major operators have indicated serious concerns about 6G on the increasingly complex hardware, along with cautiousness on the potential 6G deployment. To mitigate the concerns, the standardization of AI-Native RAN must strive for hardware reusability. When a later release of standards is available, the hardware installed for the first release should be fully utilized to host the upgraded software functions. Meanwhile, the 6G AI-Native RAN architecture and interfaces shall support centralized resource management and distributed collaborative processing for AI/ML tasks. This paradigm will not only reduce the average investment per RAN site by resource pooling, but also maximize the utilization of earlier investments by aggregating the distributed AI computing resources to support AI/ML functionalities introduced in later releases.

From the software perspective, the forward compatibility for the AI/ML models may require additional considerations beyond the traditional forward compatibility in 5G protocol designs. Due to the lack of theoretical explainability, a complex AI/ML model (e.g., DNN) may not be capable of superior performance in all conditions, and hence the RAN nodes may need to replace it with an updated version that can be of a different model structure or retrained with new AI/ML data. To that end, the 6G RAN standards shall be sufficiently flexible to enable forward compatibility for the AI/ML models. Besides, it is envisioned that AI/ML models may evolve from replacing a single processing to replacing multiple processing blocks, and eventually replacing a major part of a transceiver \cite{hoydis2021}. To enable such a dynamic application of AI/ML models, clear interface definitions are essential between the processing blocks as well as the protocol layers.

\subsubsection{\textbf{To unlock the full commercial potential of AI investments, standards should enable RAN AI service exposure}}
Different from traditional RAN functions and equipment, AI computing tends to be more expensive and energy-consuming, and therefore, its business opportunities shall be fully exploited. On one hand, AI for RAN is highly promising in automating and optimizing the networks. The improved network efficiency will naturally be the first source of commercial value from AI. For that reason, the standardization process should be strict in the identification of killer use cases for the operators, accompanied by realistic and meaningful validations to justify the gains. On the other hand, new business opportunities may exist from a resource utilization perspective. Note that AI model training is compute-intensive, but is not likely to be performed for RAN all the time. It is therefore desirable to enhance the utilization of the temporally and geographically distributed idle AI computing capacity. One approach could be similar to the cloud computing paradigm, where the RAN offers its idle AI computing capacity to AI service applications, which would create a new source of commercial value for operators beyond connectivity. For that reason, the 6G standards should also enable the operators to expose their RAN-side AI capabilities to AI service applications in an open ecosystem.

\subsubsection{\textbf{3GPP shall specify the core standards for 6G AI-Native RAN, but the success also hinges on collaboration across SDOs}}
The complexity of 6G RAN standards will go beyond the traditional RAN domain. The confluence with AI/ML and cloud technologies will have impacts not only on functional architecture and interfaces, but also on the implementations that are beyond 3GPP’s scope. To guarantee the on-time and high-quality delivery of the standards, we need the collaboration of all stakeholders.  For example, ETSI has defined network functions virtualization (NFV) and multi-access edge computing (MEC), and the O-RAN Alliance has been dedicated to RAN cloud as well as the RAN Cloud management and orchestration.

It remains imperative for 3GPP RAN to take the overarching leadership, in terms of specifying the overall functional architecture for 6G RAN and ensuring the essential interoperability (especially with the core network and UE). Meanwhile, the technical realization of 6G RAN nodes should fully consider reusing and introducing the open standards from specialized domains as well as the de-facto standards for hardware/software implementations. Towards the AI-Native 6G RAN, the SDOs need to synchronize their timeline and align their division of work properly to ensure a single and consistent set of standards.

\subsection{Key designs of AI-Native RAN for 6G Day 1}
The actual standardization of 6G AI-Native RAN will involve the use cases, the architecture, and the elementary building blocks. The whole process for AI-Native RAN standardization will follow the use case driven approach, starting with a careful identification of key use cases among those in Section III.A. The proposed functional architecture and the essential building blocks for 6G Day 1 are as follows. 

\begin{figure*} 
\centering
	\includegraphics[width=0.8\textwidth]{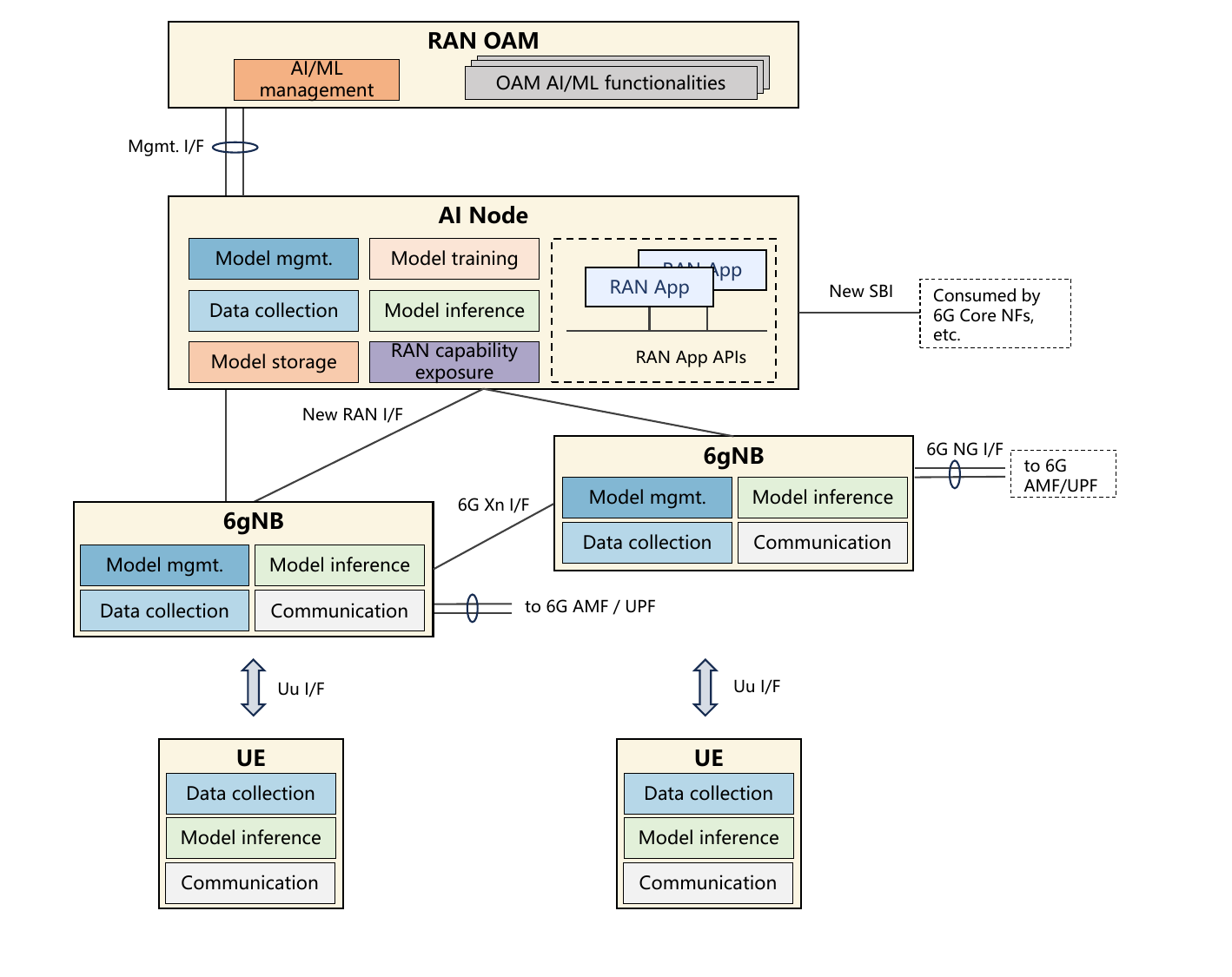}
\caption{The proposed 6G AI-Native RAN architecture.}
\label{arch}
\end{figure*}

\subsubsection{\textbf{Functional architecture}}
The proposed AI-Native RAN architecture is illustrated in Fig. \ref{arch}. The key component of the proposed architecture is a novel AI/ML-oriented RAN node, termed as “AI Node”. The AI Node and the 6G BS (“6gNB”), with a one-to-many cardinality and interconnected by a new RAN interface, constitute the 6G RAN together. The non-real-time management aspects are supported by the RAN OAM in the central cloud, which has not only its own AI/ML capabilities (e.g., management data analytics (MDA) for 6G) but also enables the MNOs with overall supervision and coordination for the AI/ML functions distributed in the RAN Nodes.  

The AI Node is equipped with sufficient AI/ML computing resources (e.g., those on an edge platform) and typically deployed as a locally centralized node. It hosts the following functionalities:
\begin{itemize}
    \item Data collection: it can collect data from RAN, UE and other domains for various AI/ML tasks including AI/ML model training, inference and performance monitoring. It may also store RAN data for future AI/ML training on demand.

    \item Model management: it can perform local management and control for the AI/ML models used for the L3 use cases, e.g., model deployment, model activation/deactivation, and switching among multiple applicable models. Moreover, it has the potential as the FL server that coordinates the distributed 6gNBs in the, if the 6gNBs are mounted with adequate AI/ML training capabilities.

    \item Model repository: it can store and catalog the AI/ML models for RAN and UE, which may be deployed at L1, L2 or L3. It may also serve as a model repository for AI service applications.
    
    \item AI/ML training: it supports AI/ML model training for RAN and UE, regardless of L1, L2 or L3 use cases. The pooled AI/ML computing resource also enables the AI Node with support for training large models for the 6G RAN. Moreover, the efficacy of model training is boosted by the digital twin capabilities in the AI Node.
    
    \item AI/ML inference: it can perform AI/ML model inference for the L3 use cases in RAN. The large models for RAN can be employed for the complex use cases as well. In addition, it may perform the AI/ML inference tasks offloaded from the 6gNBs in the collaborative inference scenarios. 
    
    \item RAN capability exposure: it is capable of exposing spare computing capacities (AI and non-AI), specific RAN data (e.g., RAN analytics), and AI/ML programmability to authorized consumers. The computing service can be consumed by OTT service applications that are deployed locally (not shown in Fig. \ref{arch}). The RAN data service may be consumed by the core network via a new service-based interface (SBI), or by OAM via the management interfaces. The AI/ML programmability can be exposed via open APIs to RAN applications that are hosted on the AI Node, which can customize the AI/ML models and associated workflow for specific use cases.
\end{itemize}

The 6gNB is the distributed RAN node for UE’s radio access. In addition to the basic communication functionalities (e.g., dynamic scheduling), it also hosts AI/ML LCM and AI/ML inference for L1/L2 use cases, as well as the associated data collection functionalities. Such an AI/ML functional split with AI/ML node ensures the real-time operations in L1/L2. This design also minimizes the AI/ML computing requirement on the 6gNBs, as the most compute-intensive AI/ML model training tasks are centralized to the AI Node. 

Some 6G UEs will also be capable of AI/ML functionalities for RAN. Nevertheless, due to its battery and form factor, such 6G UEs are likely to support AI/ML inference and associated data collection only, apart from the basic communication functionalities. 

The overall RAN architecture lays the foundation for an AI-Native RAN in 6G with the following key features:
\begin{itemize}
    \item \textbf{Efficient utilization of AI computing resource}: By offloading the sporadic AI/ML training tasks from the 6gNBs and UEs to the centralized AI Node, statistical multiplexing can be achieved. Moreover, the AI Node can offer AI/ML inference service for the 5G/6G BS without AI computing capabilities, which is especially attractive for low-cost roll-out in the developing countries. Besides, the AI computing resource may be shared with local AI service applications, and hence further enhancing the resource utilization. 
    
    \item \textbf{Hierarchical AI/ML collaboration}: The proposed design assigns the AI/ML functionalities in a systematic approach, based on the nature of diverse AI/ML tasks in RAN and their timeliness requirements. The cross-domain AI/ML collaborations with OAM or core network are enabled as well. The design also facilitates advanced learning framework like federated learning, for example, among BS and multiple UEs, or among the OAM and multiple AI Nodes.
    
    \item \textbf{Programmable AI/ML workflow}: The light-weighted RAN applications enable high programmability for the AI-Native RAN, allowing operators to customize and optimize AI/ML functionalities based on specific use cases and deployment scenarios. This paradigm ensures the 6G RAN to adapt to diversified requirements (e.g., the verticals) in an agile manner compared with the traditional device-oriented R\&D process.
\end{itemize}

\subsubsection{\textbf{Building blocks for AI-Native RAN on 6G Day 1}}
Based on the general considerations and the proposed RAN architecture, several items should be standardized with detailed technical realizations (aka “stage 3” in 3GPP and many other SDOs) for 6G Day 1, described as follows and summarized in Table \ref{Building blocks summary}:
\begin{itemize}
    \item \textbf{Data collection}: Native design for data collection is the initial step toward AI-Native RAN in 6G. The new data collection mechanisms should be introduced first for the air interface, and then on the cross-domain interfaces on 6G Day 1, as discussed in Section \ref{data collection mechanisms}. Such mechanisms involve mainly the protocols and procedures to configure data collection tasks, and how the data packets are delivered among the entities, with considerations for UE state, mobility, and coordination on basic connectivity services. The fine-grained data collection mechanisms that can be customized to diversified data collection tasks, should be continuously improved in the life cycle of 6G RAN. 

    \item \textbf{Model management}: It is well recognized that model management in 6G RAN should evolve toward an in-network and continuous AI/ML workflow, and accordingly, the model management procedures that considers the model's training and pre-validation phase before deployment and update in its life cycle, are dispensable in the Day 1 standards. Such procedures may involve capability discovery, task configuration, activation/deactivation, and status reporting, etc. Meanwhile, a more systematic set of performance monitoring designs is essential on 6G Day 1, including the specifications of the performance metric types that are applicable for the multiple collaborating models across different nodes/layers as well as the individual models. To facilitate RAN's evolution with cutting-edge AI technologies, the Day 1 standards should offer the scalability and flexibility for advanced techniques, such as FL and large foundation models. 

    \item \textbf{Collaborative AI computing}: The basic protocol support for the collaboration among the AI Node and the 6gNB should be in place on Day 1, to make the best of the proposed RAN architecture in terms of AI computing resource utilization. Such basic support would relate to the interactions between the AI Node and the 6gNB for collaborative model training and inference. The management and orchestration of the AI computing resources are also essential for enabling infrastructure reusability and ensuring forward compatibility. This requires the awareness, virtualization, monitoring, and optimized scheduling of the RAN AI infrastructure. In a later phase, more collaborative patterns, e.g., collaborative pre-training and fine-tuning, can be introduced as more AI computing capabilities are mounted in the 6gNBs. 

    \item \textbf{AI service assurance}: It is crucial for the 6G AI-Native RAN to deliver unparalleled connectivity for the emerging mobile AI services, and therefore, novel mechanisms for QoS assurance shall be available on 6G Day 1. To meet the new service requirements, the new mechanisms may involve real-time acquisition of AI service traffic characteristics and dynamic adaptation in RAN, in addition to the QoS framework enhancements that consider new service KPIs beyond the traditional ones, such as throughput, delay and packet loss. 
    When the 6G RAN evolves with new computing services beyond connectivity, the joint optimization of radio and computing resources should be considered to ensure the user experience.

    \item \textbf{RAN data service}: First, a unified service exposure framework is essential on Day 1 for RAN to expose its data services to the external consumers. The framework should enable service discovery, access control and monitoring, charging, etc. The detailed design for the service framework could be easier to start from the existing Common API Framework (CAPIF) in 3GPP \cite{3gpp.23.222}, but also needs to address the characteristics in RAN like distributed deployment and UE anonymity. On top of the service exposure framework, a number of RAN data services could come at 6G Day 1, and followed by enhanced capabilities like data augmentation at a later phase.

    \item \textbf{RAN AI computing service}: The same service exposure framework for RAN data services should be reused to expose its AI computing services. With the unified service exposure framework, it is expected that the AI computing services can be available on 6G Day 1 for both OTT service providers and RAN applications. When such services are consumed by OTT service providers, support for local offloading of the service traffic is essential in the standards. When the services are consumed by the RAN applications, standardized APIs should be exposed to enable the RAN applications to influence RAN behavior. As 6G RAN evolves, more advanced AI computing services may be considered, e.g., using RAN data (e.g., sensing data) to optimize/specialize AI models for OTT services.
\end{itemize}

\begin{table*}[]
\renewcommand{\arraystretch}{1.2}
\centering
\caption{Building blocks for AI Native RAN on 6G Day 1, and further enhancements for evolution}
\label{Building blocks summary}
\resizebox{\textwidth}{!}{%
\begin{tabular}{|l|l|l|l|}
\hline
\rowcolor[HTML]{EFEFEF} 
Essential capabilities & Standardization impact & Building blocks for 6G Day 1 & \begin{tabular}[c]{@{}l@{}} Further enhancements \\ for evolution \end{tabular}\\ \hline
\begin{tabular}[c]{@{}l@{}}AI-driven RAN \\ Processing, Optimization, \\ and Automation\end{tabular} & AI for RAN use cases & \begin{tabular}[c]{@{}l@{}}- 5G-A AI-enabled use cases;\\ - 6G new uses (e.g., AI-based RS overhead reduction),\\ considering complexity and performance gains\end{tabular} & Cross-domain use cases \\ \hline
 & Data collection & \begin{tabular}[c]{@{}l@{}}- Dedicated AI radio bearer on air interface;\\ - Cross-domain data collection\end{tabular} & \begin{tabular}[c]{@{}l@{}}Task-driven customizable \\ RAN data collection\end{tabular} \\ \cline{2-4} 
 & Model management & \begin{tabular}[c]{@{}l@{}} In network and continuous AI/ML workflow \\that enables forward compatibility \\
 - model training/retraining, inference;\\
 - model validation, model deployment and update;\\
 - model performance monitoring;\\
 - capability discovery, task configuration,\\ model activation/deactivation, and status reporting\\ 
\end{tabular} & \begin{tabular}[c]{@{}l@{}}Advanced AI/ML \\ technology support \end{tabular} \\ \cline{2-4} 
\multirow{-3}{*}{\begin{tabular}[c]{@{}l@{}}Efficient, controllable and \\ highly reliable AI LCM\end{tabular}} & \begin{tabular}[c]{@{}l@{}}Collaborative AI\\ computing\end{tabular} & \begin{tabular}[c]{@{}l@{}}- Collaborative training and Collaborative inference;\\
(for efficient AI computing resource utilization \\ and improved performance);\\ - AI computing resource management and orchestration
\\(to enable RAN infrastructure sharing, reusability, \\and forward compatibility)\end{tabular} & \begin{tabular}[c]{@{}l@{}}Collaborative pre-training \\ and fine-tuning\end{tabular} \\ \hline
 & AI service assurance & - Connection QoS assurance & \begin{tabular}[c]{@{}l@{}}Assurance for integrated\\  communication and\\ computing services\end{tabular} \\ \cline{2-4} 
 & RAN data service & \begin{tabular}[c]{@{}l@{}}- RAN service exposure framework;\\ - RAN data exposure services\end{tabular} & RAN data augmentation \\ \cline{2-4} 
\multirow{-3}{*}{AIaaS provisioning} & \begin{tabular}[c]{@{}l@{}}RAN AI computing \\ service\end{tabular} & \begin{tabular}[c]{@{}l@{}}- RAN service exposure framework (same as above);\\ - RAN AI computing services\end{tabular} &\begin{tabular}[c]{@{}l@{}} Advanced AI computing\\ services \end{tabular}\\ \hline
\end{tabular}%
}
\end{table*}

\subsection{Collaborations among 3GPP, ITU and other SDOs}

A clean split of work among 3GPP, ITU and other SDOs is crucial for the multi-SDO collaborations to fully achieve the AI-Native RAN vision. 

\begin{figure} 
\centering
	\includegraphics[scale=0.51]{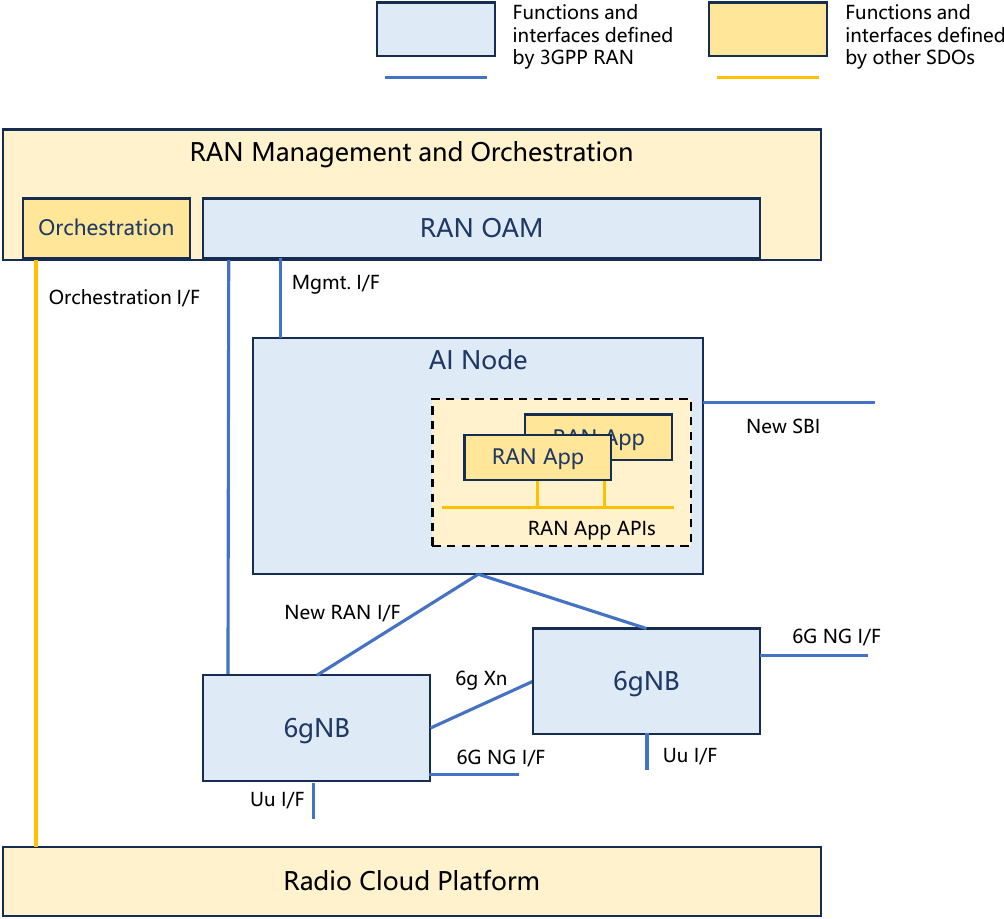}
\caption{Potential split of work between 3GPP and O-RAN, ETSI, etc.}
\label{sdos}
\end{figure}

As the link between global governments and the telecom industry, ITU has a visionary role in setting the objectives of the next-generation network. Moreover, it should take the leadership in delineating the globally-unified, umbrella framework and terminologies for AI Native networks. A relevant activity is the new Focus Group on AI-Native for Telecommunication Networks (FG-AINN) established in 2024, which aims to specify the terminologies, identify new AI/ML use cases and standardization impacts, and reshape the next-generation network architecture with native AI/ML designs \cite{FG-AINN}. 

3GPP, as the most important SDO for the telecom industry, should be responsible for the end-to-end 6G architecture, protocols, and interfaces, ensuring interoperability, scalability, and backward compatibility in commercial networks. It collaborates with industry stakeholders, regulators, and academia to address emerging 6G requirements including the integration with AI/ML, from ITU and other sources. In particular, 3GPP shall ensure multi-vendor interoperability in RAN among the base stations, and with the core network as well as the UEs.

As NFV and cloud-native paradigm begin to change how the RAN devices are implemented, other SDOs including ETSI and O-RAN Alliance may play a critical role in the RAN cloud infrastructure that host the RAN functions, which is instrumental for the forward compatibility in terms of hardware. Moreover, the SDOs other than 3GPP may also complement the RAN architecture with cutting-edge innovations, on the premise that the enhancements are compliant with the 3GPP standards, and that the work is open and non-discriminative. 

Under the umbrella framework set by ITU, different options for the split of work between 3GPP and other SDOs can be considered for the standardization of 6G AI Native RAN. A recommended option is illustrated in Fig.~\ref{sdos}. In this option, 3GPP is responsible for the 6G RAN nodes (6gNB and the new AI Node) and their management aspects, as well as the core network and the UE. The interfaces associated with those functions are also within the 3GPP scope. Apart from the hierarchical architecture associated with the AI Node, a consistent set of AI/ML LCM mechanisms shall be specified in 3GPP, which will impact all the interfaces and their protocols, in particular, the air interface and the new interface between AI Node and 6gNBs. As complement to 3GPP RAN, the RAN applications hosted by the AI Node, and the associated APIs, may benefit from the expertise in O-RAN Alliance and may continue their journey outside 3GPP. Such RAN applications can also have their variants in RAN OAM, to enable flexibility and innovation in the management domain. This paradigm has been proved valuable in 5G O-RAN with its featured hierarchy of Non-RT and Near-RT RICs. ETSI and O-RAN will also contribute to the implementation of cloud-native RAN. The RAN cloud infrastructure and its interface towards the integrated management and orchestration system will fall in scope for the SDOs other than 3GPP. It is anticipated that an open and coherent ecosystem will prosper in 6G with its promise for new services and verticals. 

An alternative option can also be considered. In this option, the AI Node, along with the hosted RAN applications, is out of scope for 3GPP but specified in O-RAN. The interfaces and protocols associated with the AI Node are also part of O-RAN's work. Apart from that, 3GPP is still responsible for the 6gNBs and their management aspects, as well as the core network and the UE. As 3GPP RAN will define the AI/ML LCM for the 6gNBs, it may be in the best interests of the telecom industry for O-RAN to align its LCM designs with those in 3GPP RAN. The collaboration among SDOs is expected to catalyze innovation in AI-native 6G RAN.

\begin{figure*} 
\centering
	\includegraphics[scale=1.2]{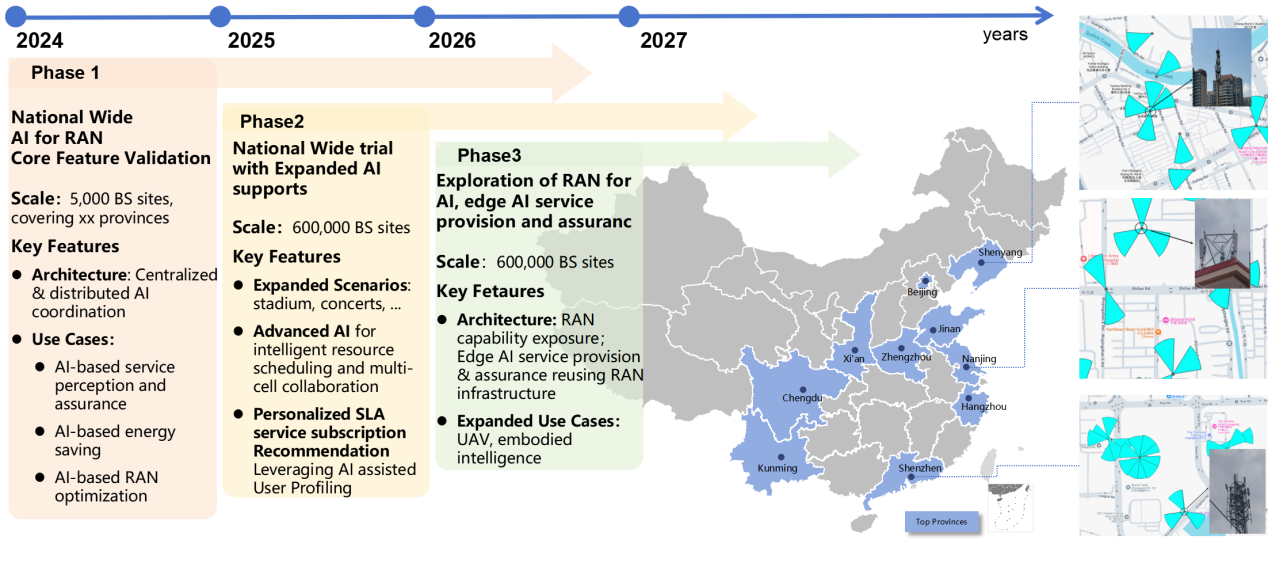}
\caption{Three phases nationwide large scale field trial.}
\label{3phase}
\end{figure*} 

\section{Nationwide Large Scale Field Trials}

\subsection{Overview}

In this section, we will share the exploration and field trial practices on the typical use cases and key features of the AI-Native RAN architecture from China Mobile’s perspective. 

To empirically validate the technical feasibility, architecture features and commercial viability of AI-Native RAN envisioned for 6G, a nationwide large-scale field trial is planned across three iterative phases as illustrated in Fig. \ref{3phase}. Built upon the nationwide 5G-A commercial network, this trial aims to systematically evaluate key features advancing the 6G vision, including the scalability of the proposed hierarchical AI-Native RAN architecture, the potential of experience-centric service models, and the convergence of communication and edge intelligence.

Phase 1 focused on the core architecture feature validation, i.e., the centralized and distributed AI coordination, and three typical use cases, including the AI-enabled differentiated service quality assurance, network energy saving, and root cause analysis for poor user experience. A centralized AI computing unit (CCU) is introduced and deployed in the trial for the computationally intensive AI inferences in the proposed solution, which reflects the AI node component as proposed in the 6G RAN architecture. Phase 1 tests started from Jan. 2024 to Dec. 2024 and covers the 31 cities in China with 5000 5G-A base stations. 

Building upon Phase 1, Phase 2 extends testing to 475,000 gNBs and incorporates diverse deployment scenarios, including high-density urban environments such as stadiums, concert venues, and tourist hotspots. This phase introduces AI-driven radio resource management (RRM) techniques, such as intelligent resource scheduling and multi-cell coordination, to enhance network performance and SLA assurance capabilities. Additionally, Phase 2 explores AI-based user profiling and proactive service subscription recommendations to monetize ``experience-centric" service models. For instance, users running extended reality (XR) applications in congested urban environments may be prompted to subscribe to an “Ultra-Low Latency” assurance plan, which offers dynamic and deterministic service guarantees that outperform those available to non-subscribed users. These advancements also validate the synergistic coordination of the RAN, network management systems (NMS), and business support systems (BSS).

Phase 3 targets nationwide deployment and addresses emerging scenarios, such as AI-driven unmanned aerial vehicle (UAV) mobility management and QoE assurance for embodied intelligence systems. Architecturally, this phase will further investigate RAN capability exposure and the feasibility of edge AI service provisioning leveraging RAN infrastructure. Phase 3 represents a pioneering step toward realizing the 6G AIaaS vision, positioning AI-Native 
RAN as foundational enablers of next-generation intelligent ecosystems.

In the following, the deployment architecture for the trials and three typical case studies in Phase 1 will be elaborated in detail.

\begin{figure*} [ht]
\centering
	\includegraphics[scale=0.6]{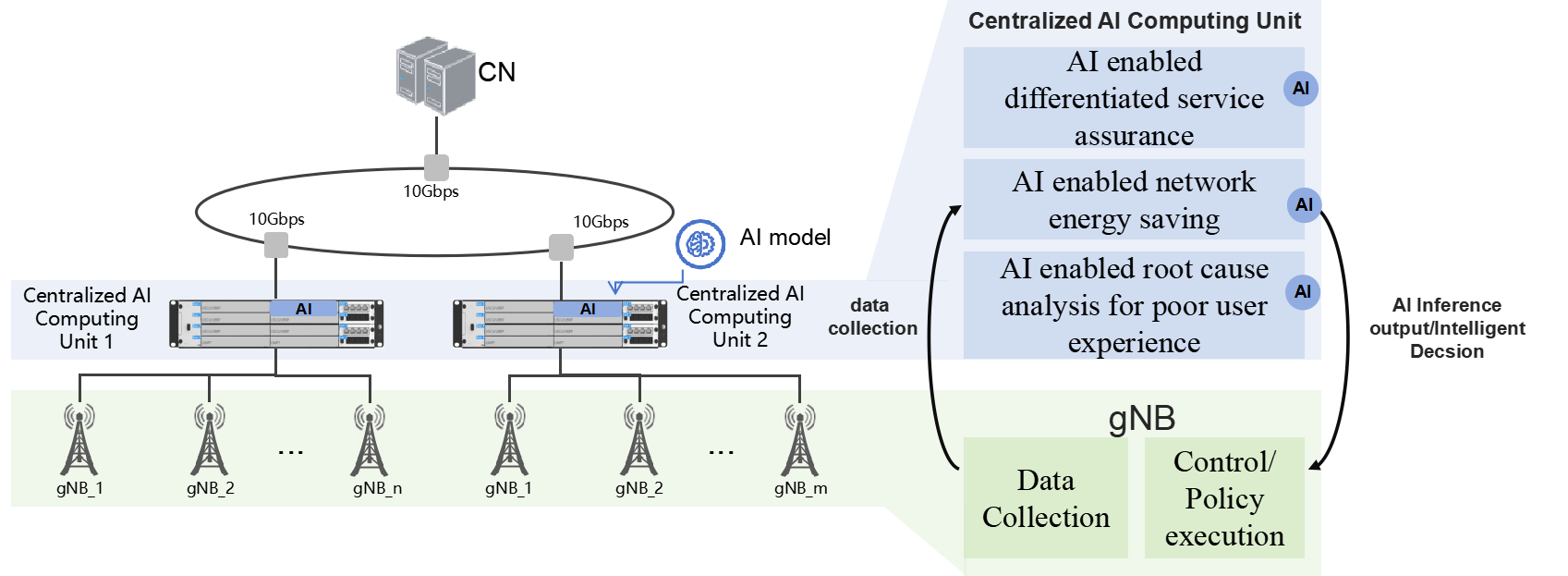}
\caption{The deployment architecture, topology and networking of the testing}
\label{topology}
\end{figure*} 

\subsection{Deployment Architecture}
The newly introduced AI-driven processes are computationally intensive, imposing demands that exceed the capabilities of existing gNBs. To resolve this limitation, a CCU—aligned with the AI node component proposed in 6G RAN architectures—was deployed in the trial. This unit consolidates computational resources to execute resource-intensive AI inference tasks across a cluster of gNBs, effectively serving as a shared AI hub. The CCU supports centralized model training, real-time inference, and cross-gNB coordination within its serviced gNB cluster. This architecture facilitates centralized and distributed AI coordination as well as elastic resource scaling, which significantly improves RAN computing infrastructure efficiency and enables dynamic resource sharing. Notably, the computing capability of the CCU can be flexibly scaled out to accommodate growing AI workloads. Furthermore, surplus computing capacity within the CCU can be repurposed to host edge AI applications (e.g., distributed video analytics), leveraging RAN infrastructure to deliver low-latency edge AI services in a cost-effective manner.

To validate the potential of AI-enabled RAN optimization, China Mobile led a large-scale field trial in 2024 with three typical use cases, shown as the Phase 1 program in Fig. \ref{3phase}. The deployment architecture, topology and networking of the testing in the field trial are illustrated in Fig. \ref{topology}.

\subsection{Case Study 1: AI-enabled differentiated service assurance}

In this subsection, we present a case study on AI-enabled differentiated service assurance. The case study also validates the superiority of the centralized AI computing unit in enabling an AI-Native RAN architecture.

To achieve deterministic service assurance for critical applications, such as those prioritized for VIP customers, an AI-empowered solution was designed. The solution employs AI-based real-time service perception to dynamically classify traffic and enforce service-level guarantees based on operator-defined policies. For example, to ensure uninterrupted short-video streaming (e.g., TikTok) for VIP users, the AI perception model identifies and monitors their TikTok traffic in real time. When service degradation is detected, QoS optimization is triggered, incorporating application-specific requirements and real-time network conditions to prioritize the VIP user’s experience. AI plays a pivotal role in this optimization process: prediction AI models can be trained to estimate user-specific metrics such as data rate, latency, and QoE, enabling proactive adjustments like priority-weighted resource scheduling and dynamic radio resource allocation. Specifically, the AI-driven optimization function can be formulated as:
\[
\hat{\mathbf{a}}_u = \arg\max_{\mathbf{a} \in \mathcal{A}} \, \mathbb{E}\left[ U\left( \hat{\mathbf{q}}_u(\mathbf{a}, \mathbf{x}_u) \right) \right],
\]
where \( \hat{\mathbf{a}}_u \) denotes the optimal action (e.g., scheduling weight, resource allocation) selected for user \( u \); \( \mathbf{x}_u \) represents the observed context of user \( u \), including application type, RAN metrics, and mobility state; \( \hat{\mathbf{q}}_u(\cdot) \) is the AI-predicted vector of service-level quality metrics (e.g., latency, throughput, QoE); \( \mathcal{A} \) is the space of feasible actions; and \( U(\cdot) \) is a utility function reflecting the operator’s policy objective, such as QoE maximization or latency minimization. This formulation allows the centralized AI model to make proactive, context-aware decisions tailored to individual users and applications.

Four representative test scenarios were selected for the trial: central business districts (CBDs), university campuses, hospital complexes, and residential zones. These scenarios encompass diverse coverage conditions (with reference signal receiving power (RSRP) ranging from -85 dBm to -115 dBm) and heterogeneous traffic load conditions, including light traffic (physical resource block (PRB) utilization $<$30\%), medium load (PRB $\in [30\%, 50\%]$), and heavy load (PRB $\in [50\%, 70\%]$).

The AI-based service perception model was trained to classify 1000+ distinct application types, achieving an accuracy exceeding 95\%. Leveraging the CCU, the system enables real-time identification of service types for 4,800 concurrent users across 8 gNBs.

\begin{figure*} 
\centering
	\includegraphics[scale=0.59]{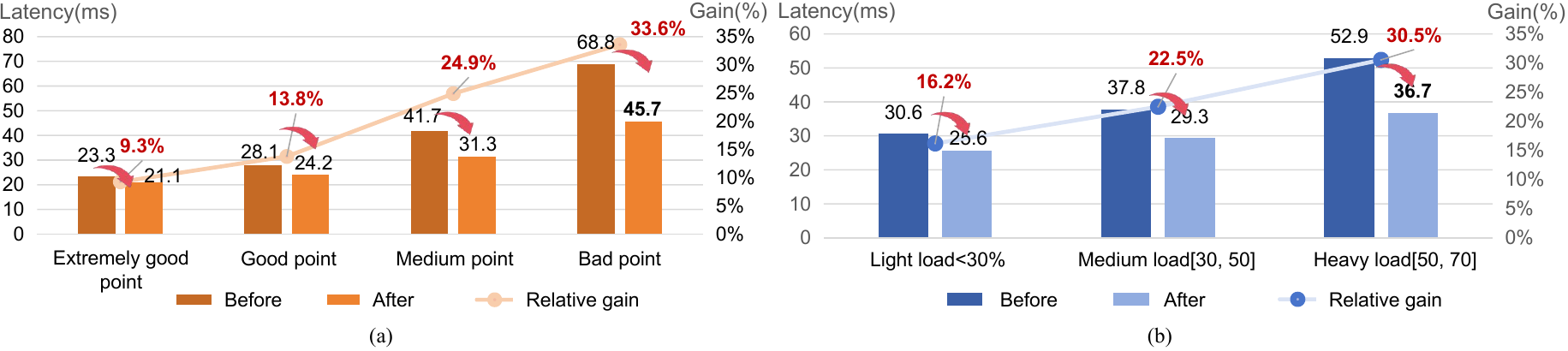}
\caption{Average air interface latency of short videos. (a) average latency (ms) vs. coverage; (b)
average latency (ms) vs. load.
}
\label{shortvideo}
\end{figure*}

\begin{figure*} 
\centering
	\includegraphics[scale=0.58]{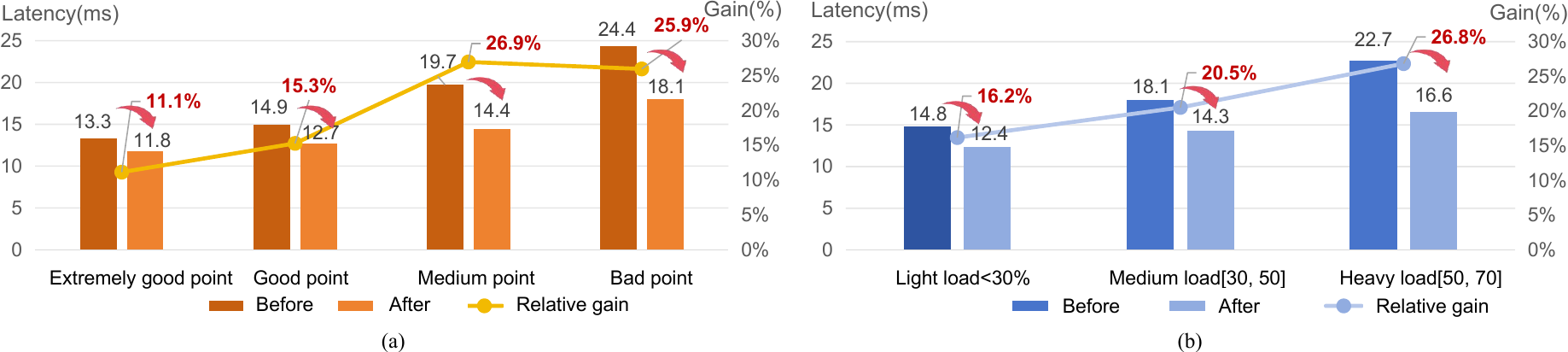}
\caption{Average air interface latency of QR code scanning application. (a) average latency vs. coverage; (b)
average latency vs. load.}
\label{QR}
\end{figure*}

To validate performance, results for two representative applications—short video streaming and QR code scanning—are analyzed. Fig. \ref{shortvideo} and \ref{QR} illustrate the latency reductions observed for these applications, respectively.


For short videos, as shown in Fig.~\ref{shortvideo}, the proposed solution significantly reduces air interface latency across diverse coverage and load conditions. In scenarios with poor coverage (RSRP $\leq$ -105 dBm) and heavy traffic loads (PRB utilization $\geq$ 50\%), latency decreased by 33.6\% and 30.5\%, respectively. The average air interface latency dropped from 43.0 ms to 32.0 ms, representing a 25.6\% improvement.

Similarly, in Fig. \ref{QR}, for QR code scanning application, latency reductions of 25.9\% and 26.8\% were achieved in scenarios with poor coverage (RSRP $\leq$ -105 dBm) and heavy traffic loads (PRB utilization $\geq$50\%), respectively. The average latency improved from 18.5 ms to 14.5 ms, yielding a 21.9\% overall reduction.

\begin{figure} 
\centering
	\includegraphics[width=0.48\textwidth]{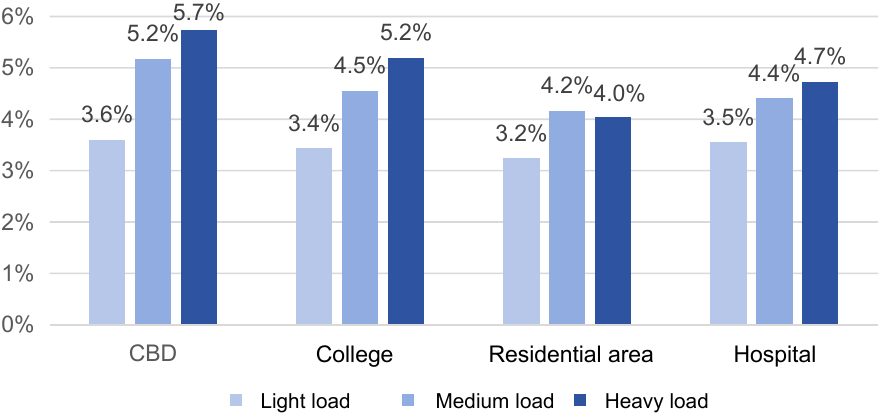}
\caption{Traffic volume growth under varying scenarios and loads }
\label{shortvideo1}
\end{figure}

It is worth noting that the proposed service assurance framework not only enhances deterministic performance but also boosts user traffic volume in 5G-A networks. Stabilized radio link quality incentivizes applications like short video platforms to dynamically upgrade streaming resolutions (e.g., from 720p to 1080p/4K), thereby improving user experience and increasing per-user data consumption
. As illustrated in Fig. \ref{shortvideo1}, the framework drives 3–6\% traffic growth across the four test scenarios, 
even under heterogeneous network conditions. 

Beyond technical performance improvements, the proposed framework enables operators to adopt experience-centric monetization strategies, shifting from traditional traffic-based models to value-added service tiers (e.g., premium ultra-low-latency subscriptions). By ensuring deterministic performance guarantees, the framework supports fine-grained service differentiation that enhances QoE and drives user adoption of premium offerings. It further positions operators to capitalize on emerging 5G-A and future 6G ecosystems by transforming network performance into a scalable revenue stream, effectively aligning infrastructure investments with long-term profitability.

\subsection{Case Study 2: AI-enabled root cause analysis for poor user experience}
\begin{figure*}
\centering
	\includegraphics[width=0.9\textwidth]{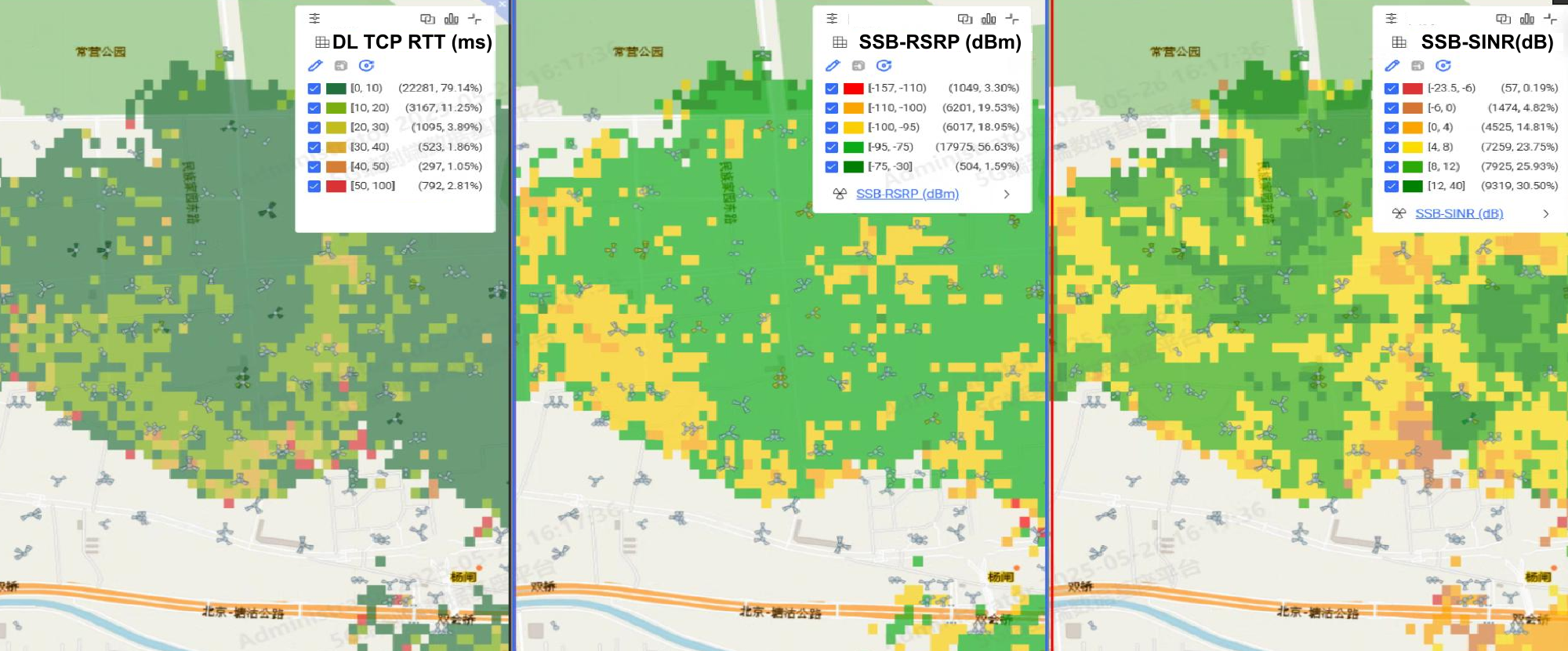}
\caption{The grid level correlated data of DL TCP RTT, SSB-RSRP and SINR. The grid size is $50m\times50m$.}
\label{uc2}
\end{figure*}

This use case targets the root causes of poor user experience by leveraging multi-dimensional data across UE, network, and application layers. A central AI unit enables fine-grained application awareness, correlating service-level metrics (e.g., throughput, delay, loss) with RAN indicators like signal to interference plus noise ratio (SINR), RSRP, handover failures, and mobility traces. These enriched per-user datasets are reported to the management system for real-time analysis. Fig. \ref{uc2} illustrates one example of the correlated downlink TCP round-trip time (RTT) data, synchronization signaling block (SSB)-RSRP and SINR in the grid granularity in the field trial, where the grid size is $50m\times50m$. 
At the management layer, multi-cell data and user session records (xDRs) are fused to enable second-level, per-user, per-service correlation. An XGBoost model classifies user types (e.g., indoor, outdoor, high-speed) with over 90\% precision, enabling tailored root cause analysis. 
To enable such high-accuracy prediction, we define a supervised learning model that maps multi-layer network observations to user-level performance labels. 
To enable such high-accuracy prediction, we define a supervised learning model that maps multi-layer network observations to user-level performance labels. The model predicts the user state or root cause category \( \hat{y}_u \) for each user \( u \) based on a fused feature vector:
\[
\hat{y}_u = f_{\theta} \left( \mathbf{x}^{\text{App}}_u, \mathbf{x}^{\text{RAN}}_u, \mathbf{x}^{\text{UE}}_u, \mathbf{x}^{\text{Mob}}_u \right).
\]

Here, \( f_{\theta}(\cdot) \) denotes the AI model (e.g., XGBoost) parameterized by \( \theta \); \( \mathbf{x}^{\text{App}}_u \) includes application-layer features such as TCP RTT and packet loss; \( \mathbf{x}^{\text{RAN}}_u \) contains radio KPIs such as SINR, RSRP, and handover statistics; \( \mathbf{x}^{\text{UE}}_u \) represents device-level attributes; and \( \mathbf{x}^{\text{Mob}}_u \) captures spatiotemporal mobility patterns (e.g., grid-based positioning). This fused modeling framework enables second-level, per-user diagnosis and supports high-precision classification across diverse radio environments.

Temporal correlation techniques further pinpoint major performance issues, such as coverage gaps, handover failures, and interference. Compared to traditional rule-based methods, this AI-driven approach improves root cause identification accuracy by 20\% and classification recall by 30\%, enabling scalable real-time diagnostics and more proactive network optimization.

\subsection{Case Study 3: AI-enabled network energy saving}
This case study demonstrates the potential of AI to revolutionize network energy management, offering a practical and scalable approach for building adaptive and sustainable green mobile networks. 

To fully tap into energy-saving (ES) potential, China Mobile has implemented more than 20 energy-saving technologies across the time, frequency, spatial, and power domains in its commercial 5G network. The core principle behind these techniques is to dynamically align network resource utilization with real-time traffic demand by switching off or scaling down cells, carriers, antennas, or PAs during periods of low traffic. To quantitatively assess the energy-saving potential of various strategies, the total energy consumption of the network is modeled as follows:
\begin{equation}
\begin{aligned}
&\mathbb{E} = \sum_{n=0}^{N} \sum_{m=0}^{M} \sum_{t=0}^{T}\\ &\left[ P^{PA}(P, m, t) + P^{\text{transceiver}}(m, t) + P^{\text{digital IF}}(m, t, C) \right] \\
&\quad + \sum_{n=0}^{N} \sum_{t=0}^{T} \left[ P^{\text{baseband}}(t, C) + P_0 \right].
\end{aligned}\nonumber
\end{equation}

In this model, $\mathbb{E}$ denotes the total energy consumption over the network. $N$, $M$, and $C$ represent the number of base stations, channels (spatial domain), and carriers (frequency domain), respectively. $T$ denotes the operating time, and $P$ is the transmit power. $P^{PA}$, $P^{\text{transceiver}}$, $P^{\text{digital IF}}$, and $P^{\text{baseband}}$ denote the power consumption of power amplifiers, transceivers, digital intermediate frequency modules, and baseband processors, respectively. $P_0$ represents the static power consumption of other essential components. This formulation allows for fine-grained analysis and comparison of energy usage across different domains and system elements. However, the above mentioned ES approaches often depend on manually configured parameters, such as fixed time windows and static traffic thresholds, which lack the agility to respond to dynamic traffic patterns, resulting in suboptimal energy efficiency.

To overcome these limitations, two AI-enabled energy-saving solutions are proposed and validated in the large-scale field trial. For solution 1, AI is integrated into the network management system to predict traffic patterns and dynamically adjust energy-saving thresholds based on cell-level data sampled at a 15-minute interval. This improves flexibility over static configurations but remains limited by coarse data granularity and slower reaction times. While for solution 2, CCUs are further deployed at base stations to enable real-time service type recognition. By identifying delay-tolerant traffic, AI enables intelligent scheduling strategies such as packet aggregation, extending radio resource sleep periods without degrading user experience. This edge-intelligent approach complements centralized management and significantly boosts energy-saving efficiency through fine-grained, service-aware optimization.

All results are benchmarked against a baseline scenario without energy-saving mechanisms. Compared to this baseline, the proposed AI-based approaches show significant improvements: solution 1 achieves 26.6\% energy saving, while solution 2 further increases the savings to 34.16\%. This represents an additional 5.32\% and 12.88\% improvement, respectively, over traditional non-AI energy-saving methods.

\begin{figure}[t]
\centering
	\includegraphics[width=0.47\textwidth]{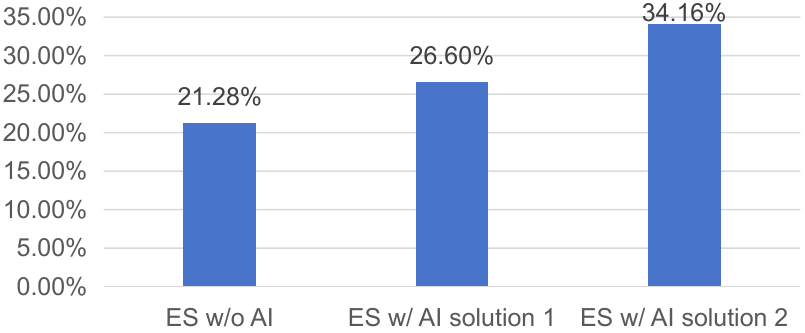}
\caption{Performance gain of the AI-enabled energy saving.}
\label{uc_ee}
\end{figure}

\subsection{Insights for 6G}
The three-phase nationwide field trial is pivotal for advancing 6G and offers critical insights into the integration of AI-Native architectures within 5G-A/6G RANs. Phase 1’s validation of the CCU coordinating multiple gNBs—provides a foundational reference for 6G standardization, demonstrating how centralized intelligence enhances resource efficiency and deterministic service assurance. However, the trials also revealed unresolved challenges: the absence of standardized interfaces between the CCU and multi-vendor gNBs hindered cross-platform collaborative inference, while the lack of modular API encapsulation for the AI capabilities limited their openness and reusability across diverse scenarios. These findings underscore the necessity for 6G standardization bodies to prioritize interoperable AI service frameworks, enabling vendor-agnostic coordination and flexible deployment of network-native intelligence.

\section{Conclusions}
This paper presents an operator-oriented view on AI-Native RAN design for 6G, with a focus on potential Day 1 standardization priorities. Building on a review of AI/ML integration in 5G and 5G-A, we first articulate the concept of AI-Native RAN and its three essential capabilities. For each capability, we identify key enabling technologies and discuss corresponding Day 1 standardization considerations.

The paper further articulated our perspective on the scope and objectives of Day 1 standardization for AI-Native RAN in 6G. In particular, we proposed that the introduction of an AI Node is essential and should be realized from Day 1. To validate the proposed architecture, concepts, and key enabling technologies, a large-scale field trial was conducted throughout 2024, spanning 31 cities in China and involving over 5000 5G-A base stations. The results demonstrated significant improvements, including a 25.6\% and 21.9\% reduction in latency for short video streaming and QR code scanning, respectively, along with a 3–6\% increase in overall network traffic across diverse scenarios. The trial highlighted the value of centralized AI in enhancing resource efficiency and service assurance, while also revealing gaps in standardized interfaces and modular APIs for AI capabilities. In addition, field trials on energy saving and root cause analysis use cases were performed. Notably, the proposed AI-driven root cause analysis framework achieved up to 20\% improvement in diagnostic accuracy and a 30\% gain in user type classification recall compared to rule-based methods, enabling scalable and proactive network optimization.

Moving forward, as the 6G standardization progresses, we anticipate further refining the design of AI-Native RAN and conducting more validations in diverse practical scenarios. It is expected that the research and practice presented in this paper will provide strong support for the standardization and commercialization of 6G AI-Native RAN, and drive the development of communication networks towards intelligence, efficiency, and sustainability.

\section{Acknowledgment}

This work was supported by the National Key Research and Development Program of China under Project 2022YFB2902100.


%


\bibliography{main}

\bibliographystyle{IEEEtran}

\end{document}